%% file: mnras_template.tex
\documentclass[fleqn,usenatbib]{mnras}

\usepackage{newtxtext,newtxmath}

\usepackage[T1]{fontenc}

\DeclareRobustCommand{\VAN}[3]{#2}
\let\VANthebibliography\thebibliography
\def\thebibliography{\DeclareRobustCommand{\VAN}[3]{##3}\VANthebibliography}

\usepackage{graphicx}	
\usepackage{amsmath}	
\usepackage{booktabs} 

\usepackage{float} 

\title[Opacity tables for giant planets]{Mean opacity tables for probing the interior and atmosphere of giant planets}

\author[L. Siebenaler and Y. Miguel]{
Louis Siebenaler$^{1}$\thanks{E-mail: siebenaler@strw.leidenuniv.nl} and 
Yamila Miguel$^{1,2}$
\\
$^{1}$Leiden Observatory, University of Leiden,
              Einsteinweg 55, 2333CA Leiden, The Netherlands\\
$^{2}$SRON Netherlands Institute for Space Research , Niels Bohrweg 4, 2333CA Leiden, The Netherlands
}

\date{Accepted 2025 December 9. Received 2025 December 4; in original form 2025 October 23}

\pubyear{\the\year{}}

\begin{document}
\label{firstpage}
\pagerange{\pageref{firstpage}--\pageref{lastpage}}
\maketitle

\begin{abstract}

We present new Rosseland and Planck mean opacity tables relevant to the shallow interiors and atmospheres of giant planets. The tables span metallicities from 0.31 to 50 times solar, temperatures from $100 - 6000 \ \rm K$, and pressures from $10^{-6} - 10^{5} \ \rm bar$, thereby covering a wider parameter space than previous data sets. Our calculations employ the latest molecular and atomic line lists and pressure-broadening treatments, and include contributions from collision-induced absorption, free electrons, and scattering processes. We further provide cloudy mean opacity tables that account for cloud particle extinction across a range of particle sizes and capture the sequential removal of condensates as the gas cools.
We benchmark our cloud-free tables against widely used opacity tables and find significant relative differences, exceeding 100\% in Rosseland mean opacities at $T \gtrsim 3000 \ \rm K$ due to the inclusion of additional short-wavelength absorbers. Differences in Planck mean opacities at high temperatures are even larger, in some cases exceeding two orders of magnitude, which is most likely driven by the inclusion of Ca, Mg, and Fe cross-sections and updated Na D and K I resonance line treatments. Cloud opacities substantially increase Rosseland mean opacities for $T \lesssim 2800 \ \rm K$, while their effect on Planck mean opacities is weaker. We also discuss limitations of our mean opacities at high pressures, where non-ideal effects become important. 
This work provides improved cloud-free mean opacity tables for giant planets, as well as the first publicly available cloudy mean opacity tables, which will enable more realistic modeling of their atmospheres and interiors.

\end{abstract}




\section{Introduction}

Radiative transport controls the thermal structure, evolution and formation of stars and planets. However, performing full radiative transfer calculations is computationally costly and often unfeasible due to the frequency dependence of opacity. A common solution is to use pre-tabulated mean opacity tables, which are frequency independent and depend only on the temperature, pressure, and composition of the medium. The most widely used are the Rosseland mean opacity ($\kappa_{\rm R}$), appropriate for optically thick, diffusive regimes, and the Planck mean opacity ($\kappa_{\rm P}$), relevant for optically thin conditions. Both are essential inputs for stellar evolution models (e.g. \citealt{Paxton_2011}), giant planet interior models (e.g. \citealt{Guillot_1994b, Sur_2024}), and analytical atmospheric models of planets (e.g. \citealt{Guillot_2010, Parmentier_2014, Heng_2014}). In these models, mean opacities will determine the thermal structure of the star or planet.

Several mean opacity tables have been developed over the past decades for astrophysical applications. In the context of Solar system and exoplanets giant planets, the most commonly used are the \cite{Freedman_2008} (hereafter F08) and \cite{Freedman_2014} (hereafter F14) tables. These data sets were based on then state-of-the-art molecular and atomic line lists, together with pressure-broadening treatments. However, the rapid growth of high quality exoplanet observations enabled by \emph{James Webb Space Telescope} (\emph{JWST}), the \emph{Very Large Telescope} (\emph{VLT}), and upcoming missions such as the \emph{Atmospheric Remote-sensing Infrared Exoplanet Large-survey} (\emph{ARIEL}) \citep{Tinetti_2022} has created an increasing need for updated opacity data to construct reliable atmospheric models. As a result, since the release of F14, significant advances have been made in improved and more extensive molecular line lists thanks in large to the efforts from the ExoMol \citep{Tennyson_2024} and HITRAN \citep{Gordon_2022} teams. Additionally, improvements on the sodium and potassium resonance lines have been made \citep{Allard_2016, Allard_2018, Allard_2023, Allard_2024, Allard_2025}, which are known to be key opacity sources in giant planets (e.g. \citealt{Guillot_2004, Siebenaler_2025}). Another limitation of the F08 and F14 tables is the absence of condensate opacities. Yet, clouds are well known to be fundamental absorbers in giant planets, and have been shown to alter their thermal structure and evolution tracks (e.g. \citealt{Poser_2019, Poser_2024, Morley_2024}). Accurate evolution models, in turn, are increasingly important for constraining planetary properties such as radius, mass, and bulk metallicity. With these points in mind, it is timely to compute new mean opacity tables that incorporate both the latest improvements in molecular and atomic cross-sections and the effect of clouds.

In this work, we compute new radiative mean opacity tables of $\kappa_{\rm R}$ and $\kappa_{\rm P}$ appropriate for the different chemistries of hydrogen-dominated atmospheres, relying on the latest available opacity data. We provide both cloud-free and combined gas–cloud tables (hereafter referred to as cloudy mean opacities), spanning a broad temperature and pressure range, $100 - 6000 \ \rm K$ and $10^{-6} - 10^{5} \ \rm bar$, thereby extending the parameter space of F08 and F14. In Section \ref{sec:methods}, we explain the method of the chemistry calculation, and the sources of opacity used in this work and how they were computed. In Section \ref{sec:results}, we present our cloud-free mean opacities and compare them to the F14 tables. We also present our cloudy mean opacities and show how the assumed cloud particle size affects the results. Section \ref{sec:discussion} demonstrates the impact of our opacity tables on an evolution model of a Jupiter-like planet, and discusses the main uncertainties related to high-pressure opacities. In Section \ref{sec:conclusion}, we give our conclusions.

\section{Methods} \label{sec:methods}

\subsection{Chemistry calculations}

\begin{figure*}
	\includegraphics[width=1\textwidth]{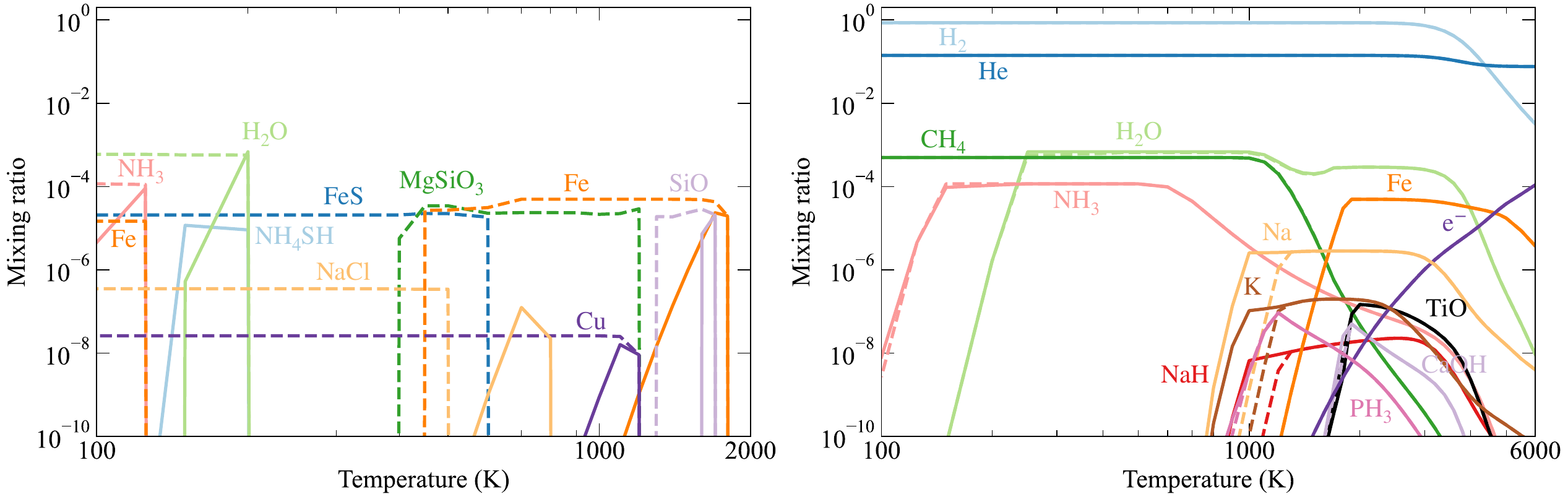}
    \caption{Number mixing ratio of species at a fixed pressure of $1 \ \rm bar$ assuming a solar composition from \citet{Asplund_2021}. The left panel shows the chemistry of condensates using the rainout approach (solid curves) and equilibrium condensation (dotted curves). The right panel is the same as the left panel but showing the gas chemistry.}
   \label{fig:chemistry_rainout_equ}
\end{figure*}

To construct the mean opacity tables, we use thermochemical equilibrium abundances computed with the code \texttt{GGchem} \citep{Woitke_2018}. This allows condensation to be treated in two ways: through equilibrium condensation or the rainout approximation.

In equilibrium condensation, the gas and solid phases remain in thermodynamic equilibrium at all temperatures and pressures. Condensates that form at high temperatures (primary condensates) stay well mixed with the gas and continue to interact with gaseous species to produce secondary condensates at lower temperatures. A well-known example of a primary condensate is Fe, which enables the formation of the secondary condensate FeS once the gas cools. This approach is typically used to model the chemistry in low-gravity environments, such as protoplanetary discs \citep{Jorge_2022, Oosterloo_2024}.

In contrast, under the rainout approximation, once a condensate forms it settles gravitationally (also referred to as rainout) and is removed from the overlying atmosphere. With no further equilibrium between the gas and solid phases, primary condensates can no longer participate in reactions at lower temperatures to form secondary condensates. This process is similar to cloud settling in high-gravity environments. Observations of Solar System and exoplanetary giant planets suggest that the rainout approximation provides a more realistic description of their chemistry.

Fig. \ref{fig:chemistry_rainout_equ} illustrates how the chemistry of condensates (left panel) and gas (right panel) changes when using the rainout or equilibrium condensation approach. In equilibrium condensation (dashed curves), Fe condensates remain in the atmosphere down to $\sim 300 \ \rm K$, where it reacts with sulfur-bearing gas to form the termochemically favourable secondary condensate FeS. In the rainout case (solid curves), Fe settles into a deep cloud layer, and its concentration is strongly reduced, preventing FeS formation. Consequently, cloud/grain opacities differ significantly between the two chemistry approaches. Previous works, such as \citet{Ferguson_2005} and \citet{Marigo_2024}, modelled grain opacities in their mean opacity tables using equilibrium condensation. To the best of our knowledge, there are currently no publicly available mean opacity tables that model cloud opacities using the rainout approximation, despite its relevance for modeling planetary atmospheres and interiors. Differences in gas chemistries between the two approaches are for the most part negligible in mean opacity calculations. The only notable impact comes from the depletion of Na and K in the gas phase, which occurs at lower temperatures in the rainout approximation, as shown in the right panel of Fig. \ref{fig:chemistry_rainout_equ}.

In this work, we focus on computing mean opacity tables using the rainout approach, given its suitability for planets \footnote{We have also computed mean opacity tables using the equilibrium condensation approach. They are briefly discussed in Appendix \ref{sec:appendix_equ_cond}.}. To approximate realistic atmospheric behaviour, we compute the rainout chemistry along isobaric profiles, starting at high temperatures and progressing towards lower temperatures. When a new isobar is considered, the chemistry is reset. This approach captures the sequential removal of condensable species as the gas cools.

\subsection{Opacity sources and calculations}

In this section, we introduce the opacity sources considered in this work and describe how they are calculated. To account for the diversity of giant planets, we compute opacities over a broad temperature range $100 - 6000$ K. Additionally, to ensure that our opacity tables are applicable to both planetary atmospheres and shallow interiors, we compute opacities across a pressure range of $10^{-6} - 10^5$ bar. In general, when possible, we compute the opacities over a spectral domain from 0.1 to 500 $\mu$m. In total, we compute wavelength-dependent opacities at 1228 pressure–temperature points on a nearly square grid.

\subsubsection{Molecular opacities}

\begin{figure*}
	\includegraphics[width=0.9\textwidth]{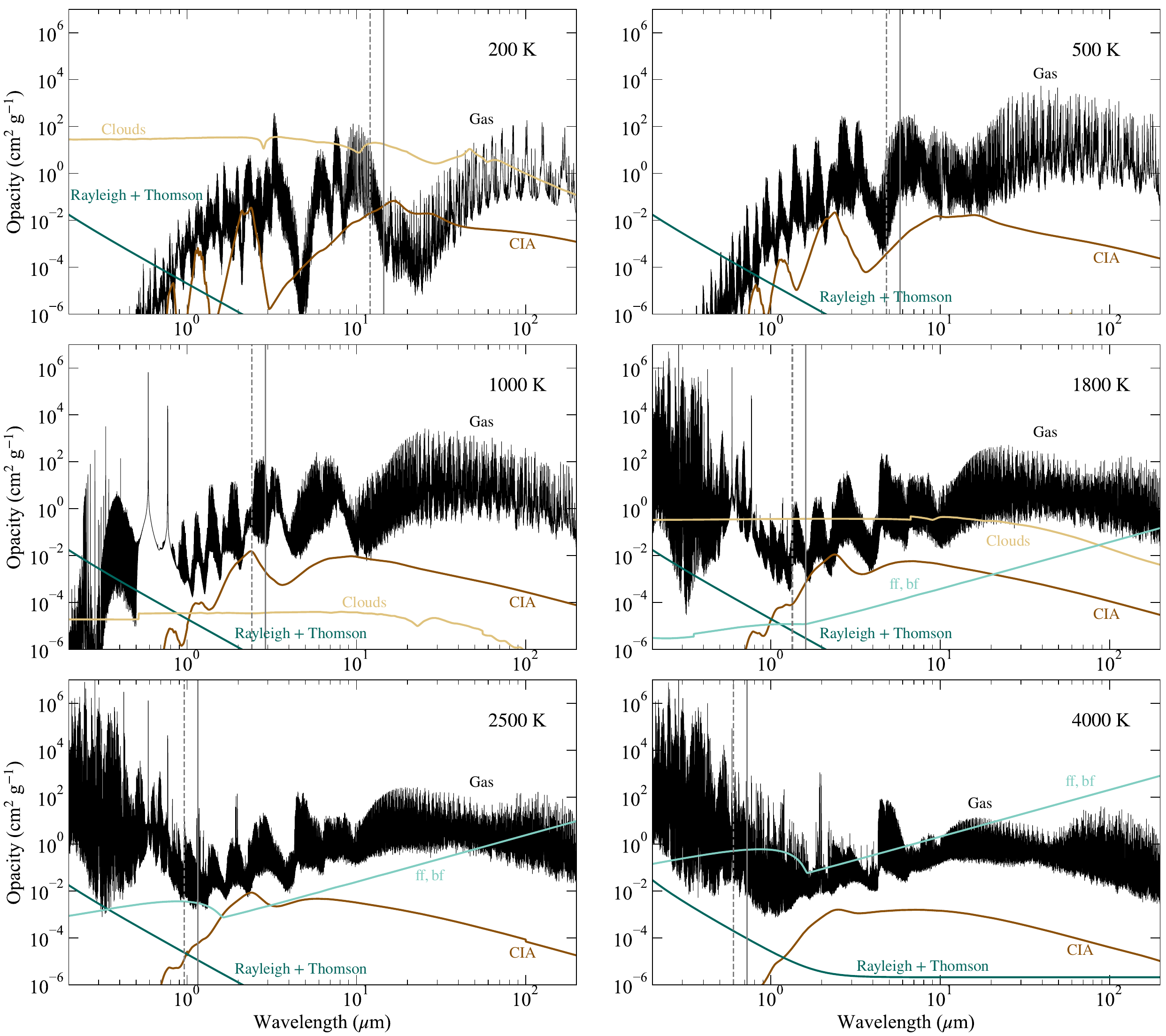}
    \caption{Monochromatic opacities at 1 bar for a solar composition. Each panel applies to a different local gas temperature $T_{\rm g}$. The opacities of all neutral molecules and atoms are shown in black, CIA is brown, free-free (ff), and bound-free (bf) absorption is turquoise, Rayleigh and Thomson scattering is given in cyan, and the opacity due to clouds is yellow. To compute the cloud opacity a mean particle radius $r_{\rm g} = 1 \ \rm \mu m$ was used. The solid and dotted grey lines mark the wavelengths where the Planck function $B_{\lambda}$ and its temperature derivative $dB_\lambda/dT$, respectively reach their maxima for the local gas temperature $T_{\rm g}$.}
   \label{fig:opacity_spectra}
\end{figure*}

The cross-section of molecules are calculated using the Python package \texttt{pyROX} \citep{De_Regt_2025}, which has been used in several previous studies \citep{deRegt_2025, Gonzales_2025, Siebenaler_2025}. This package computes the line strengths and broadening widths for individual transitions based on molecular line lists, partition functions, and broadening parameters. Absorption lines are modelled using Voigt profiles, expressed as the real part of the normalized Faddeeva function \citep{Gandhi_2020}. Table \ref{table:molecule_opacity} summarizes the molecules considered in this study, along with their corresponding line lists.

The line broadening of the Voigt profile requires broadening formalisms of the Gaussian and Lorentzian component. We describe the Gaussian profile with a half width at half maximum (HWHM; in cm$^{-1}$; \citealt{Hill_2013})

\begin{equation}
    \gamma_{\rm G} = \frac{\nu_0}{\rm c} \sqrt{\frac{2\textrm{ln(2)k}_{\rm B}T}{m}},
\end{equation}

where $\nu_0$ is the line centre in wavenumber and $m$ is the mass of the specie. For the Lorentzian component, we adopt the ExoMol \citep{Tennyson_2016, Tennyson_2024} formalism, which computes the  HWHM (in cm$^{-1}$ atm$^{-1}$) as 

\begin{equation}
    \gamma_{\rm L} = \gamma_{\rm N} + \sum_b \gamma_{0,b} \bigg(\frac{296 \ \rm K}{T}\bigg)^{n_b} \frac{P_{b}}{1 \ \rm atm},
\end{equation}\label{eq:gamma_L_molecules}

where $\gamma_{0,b}$ is the Lorentz coefficient (in cm$^{-1}$ atm$^{-1}$) for a line broadened by collision with a specie $b$, and $n_b$ describes its temperature dependence. The partial pressure of the perturber $b$ is given by $P_{b}$ (in atm), and is computed using the ideal gas law. In this work, we consider H$_2$ and He as perturbers, assuming a background atmosphere composed of $85\%$ H$_2$ and $15\%$ He, typical of Jupiter-like planets. $\gamma_{\rm N}$ is the natural broadening coefficient (in cm$^{-1}$ atm$^{-1}$) and has a less significant effect. The broadening parameters $\gamma_{0,b}$ and $n_b$ depend on the potential energy curve describing the interaction between the perturber and molecule, and as such are specie and transition dependent. Typically, these parameters are determined through experiments or ab-initio calculations. For some molecules, the ExoMol database provides detailed H$_2$ and He pressure-broadening coefficients in terms of the energy state quantum number $J$. We make use of this data for the following molecules: H$_2$O \citep{Solodov_2008, Solodov_2009, Voronin_2010, Petrova_2013, Petrova_2016, Barton_2017}, CH$_4$ \citep{Varanasi_1972, Varanasi_1989, Varanasi_1990, Pine_1992, Fox_1998, Grigoriev_2001, Gabard_2004, Fissiaux_2014, Lyulin_2014, Manne_2017, Vispoel_2019}, NH$_3$ \citep{Guest_2024}, PH$_3$ \citep{Sergent_Rozey_1988, Levy_1993, Kleiner_2003, Salem_2004, Salem_2005}, CO \citep{Gordon_2017, Guest_2024}, and HCl \citep{Wilzewski_2016}. For the molecules AlH, CaH, MgH, CrH, FeH, TiH, SiO, TiO, and VO, we make use of the $J$-dependent broadening coefficients estimated by \cite{Gharib-Nezhad_2021}. These are not based on experimental data or ab initio calculations, but instead rely on a $J$-dependent collision theory extrapolated from CO and HCl broadening data. For the remaining molecules, we define the broadening parameters based on the approach from the ExoMolOP database \citep{Chubb_2020}. In this method, a molecule with known broadening parameters is identified based on a similar dipole moment, molar mass, and general structure (linear, non-linear, diatomic, polar, non-polar) to the target molecule, and the mean values of $\gamma_{0,b}$ and $n_b$ from the reference molecule are adopted.

While it is common practice to compute absorption cross-sections using a Voigt profile, it is known from spectroscopic measurements that non-Lorentzian behaviour can be important in both the line-centre and wings \citep{Hartmann_2002, Ngo_2013}. Including non-Lorentz behaviour in the calculation of opacities is challenging due to a lack of spectroscopic parameters and a first principle theory. Hence, to minimize the uncertainties related to the line wing, one introduces a line wing cut-off $R_{\rm cut}$, which defines the extent of the line wing on either side from the line centre. In this work, we adopt the proposed standard practice procedure from \cite{Gharib_Nezhad_2023},

\begin{equation}
    R_{\rm cut} = 
    \begin{cases}
        25 \ \rm cm^{-1} & \textrm{for } {P} \leq 200 \textrm{ bar} \\
        100 \ \rm cm^{-1} & \textrm{for } {P} > 200 \textrm{ bar}
    \end{cases}
\end{equation}

After employing the line wing-cut-off, we renormalize the line profile to ensure that the integrated line strength is conserved (see equation A6 in \citealt{Lacy_2023}).

Several state-of-the-art line lists contain billions of absorption lines, making the computation of opacities extremely intensive. To improve computational efficiency, we apply a super-lines method for selected molecules, similar to the approach used in the \texttt{ExoCross} code \citep{Yurchenko_2018} and adopted in the ExoMolOP database.
In this method, we define a local cut-off parameter $s$, which is used to identify “weak” lines within a narrow wavenumber bin. Lines that contribute less than a fraction $s$ to the total line strength within the bin are considered weak. Their combined strength is then added to the strongest line in the bin, thus producing a super-line. This allows us to omit the individual Voigt profiles of weak lines from the opacity calculation, while preserving the total integrated line strength within each bin. We employ this method using $s=0.35$ \citep{deRegt_2025} and wavenumber bins of size 0.001 cm$^{-1}$ to the following molecules: H$_2$O, CH$_4$, NH$_3$, PH$_3$, VO, CO$_2$, LiOH, H$_2$S, and CaOH. 

\subsubsection{Atomic opacities}
Atomic opacities are also computed using \texttt{pyROX}, based on data from NIST \citep{NIST_2001}, the Vienna Atomic Line Database (VALD; \citealt{Ryabchikova_2017}) and the Kurucz database \citep{Kurucz_2018}. We consider the following neutral species: Ca, Cr, Fe, K, Li, Mg, Mn, Na, Ni, Ti, V. Line lists for all atoms were taken from VALD, except for Fe, where the Kurucz data were used. For details on the line broadening formalism, we refer the reader to section 2.3.1 of \cite{Siebenaler_2025}. 

The Na D and K I resonance lines are modelled using the data from \cite{Allard_2016}, \cite{Allard_2019}, and \cite{Allard_2025} which account for perturbations by H$_2$. These calculations are valid upto H$_2$ number densities of $N_{\rm H_2} = 10^{21} \ \rm cm^{-3}$. Beyond this critical density, we model the Na D and K I resonance lines using Voigt profiles where the HWHM is computed using the impact approximation with broadening parameters from VALD, and apply a line wing cut-off of $R_{\rm cut} = 4500 \ \rm cm^{-1}$, similar to \cite{Siebenaler_2025}.

\subsubsection{Collision-induced absorption opacities}

Collision-induced absorption (CIA) arises during the close encounter of two interacting species, which induces a transient dipole moment that enables rototranslational (RT) and rotovibrational (RV) transitions. This produces a continuum opacity rather than distinct spectral lines, and has long been known to play a significant role in Solar system giant planets \citep{Trafton_1967}. Table \ref{table:CIA_opacity} summarizes the CIA opacity sources included in this work. For most collision pairs we use a single data source, with the exception of H$_2$–H$_2$ and H$_2$–He, for which multiple data sources are combined to cover different temperature and wavelength ranges. 

For H$_2$–H$_2$ CIA below 400 K, we adopt the RT spectra from \cite{Fletcher_2018} and \cite{Orton_2025} for $\lambda > 2.5 \ \rm \mu m$, while the RV spectra from \cite{Borysow_2002} are used at shorter $\lambda$. Between 400 and 3000 K, we use the RT and RV spectra from \cite{Abel_2012} for $\lambda > 1 \ \rm \mu m$ and supplement them with RV data from \cite{Borysow_2001} and \cite{Borysow_2002} at shorter $\lambda$. For temperatures above 3000 K, we use the data from \cite{Borysow_2001}.

For H$_2$–He CIA below 200 K, we use the RT spectra from \cite{Orton_2025} at $\lambda > 4.17 \ \rm \mu m$ and the RV data from \cite{Borysow_1989} and \cite{Borysow_1989_2} at shorter $\lambda$. At higher temperatures, the data from \citet{Abel_2011} is used.




\subsubsection{Cloud opacities}\label{sec:grain_opacities}
To compute the opacity of cloud particles, we require both their absorption and scattering properties, as well as their abundances as a function of temperature and pressure. At a given temperature and pressure, we compute the monochromatic opacity (in $\rm cm^2 \ g^{-1}$) due a given cloud particle as

\begin{equation}
    \kappa_{\lambda, \rm cloud} = \frac{\int n(r, r_{\rm g}, \sigma_{\rm g}) Q_{\rm ext}(r, \lambda) dr}{\rho_{\rm atm}}, 
\end{equation}

where $Q_{\rm ext}$ (in $\rm cm^2$) corresponds to the extinction efficiency (absorption + scattering) of the cloud particle, which depends on its radius $r$ and wavelength $\lambda$. We compute $Q_{\rm ext}$ using the code \texttt{LX-MIE} \citep{Kitzmann_2017}, based on Mie theory. The atmospheric mass density is denoted $\rho_{\rm atm}$, and $n$ describes the particle size distribution, for which we adopt a lognormal distribution, similar to \cite{Ackerman_2001}. We have

\begin{equation}
    n(r, r_{\rm g}, \sigma_{\rm g}) = \frac{N_{\rm cloud}}{r\sqrt{2\pi}\rm ln\sigma_{\rm g}}\textrm{exp}\bigg[-\frac{\textrm{ln}^2(r/r_g)}{2\rm ln^2\sigma_g}
    \bigg],
\end{equation}

with a standard deviation $\sigma_{\rm g}$ and a particle mean radius $r_{\rm g}$. The total number density of cloud particles is then calculated as 

\begin{equation}
    N_{\rm cloud} = \frac{3\rho_{\rm atm} \varepsilon q_{\rm cond}}{4\pi \rho_{\rm cloud}r_{\rm g}^3} \textrm{exp}\bigg(-\frac{9}{2} \rm ln^2 \sigma_g  \bigg),
\end{equation}

where $\varepsilon$ is the ratio of condensate to atmospheric molecular weight, $q_{\rm cond}$ is the number mixing ratio of the condensate, and $\rho_{\rm cloud}$ is the mass density of the condensed particle. 

A detailed description of $q_{\rm cond}$ and $r_{\rm g}$ would require modeling physical processes such as condensation, sedimentation, and eddy mixing. 
However, incorporating all these processes is not feasible on chemistry grids, which are constructed along isobaric temperature profiles and lack an underlying atmospheric structure. Instead, we adopt a simplified approach, where we use the output of the chemistry model \texttt{GGChem} for $q_{\rm cond}$, thereby neglecting the effects of vertical mixing. This results in more confined cloud layers compared to models that include mixing, but still provides a first-order approximation of where clouds influence the opacity. We also neglect sedimentation and treat $r_{\rm g}$ as a free parameter. Given the large diversity in cloud partile sizes in planetary atmospheres \citep{Ormel_2019, Ohno_2020, Huang_2024}, this assumption is reasonable. In this work, we construct mean opacity tables with $\sigma_{\rm g} = 2$ and for a range of $r_{\rm g}$ values ($0.01$, $0.05$, $0.1$, $0.5$, $1$, $2$, $5$, $10$, $50$ $\mu$m) allowing users to interpolate between them based on their best estimate of the relevant cloud particle sizes in a given layer. While simplified, this framework represents an improvement over commonly used parametric prescriptions of cloud decks in planetary evolution models (e.g. \citealt{Heng_2012, Poser_2024}). By explicitly tabulating opacities for a wide range of particle sizes and using optical data of condensates, our tables enable a more physically grounded treatment of clouds across diverse planetary conditions.

Table \ref{table:cloud_opacity} summarizes the cloud particles that were considered in this study along with the sources of their optical constants. Nitrogen bearing species, including NH$_4$SH and NH$_3$, are expected to condense in cold giant planets like Jupiter. \cite{Howett_2006} measured optical properties of NH$_4$SH ice between $2.5-7.7 \ \rm \mu m$. However, NH$_4$SH condenses at $\sim200$ K on Jupiter, where the Planck function peaks near $15 \ \mu \rm m$. To extend the spectral coverage, we supplement this with optical constants for NH$_4$CN from \citep{Gerakines_2024} from $2 - 2.5 \ \mu \rm m$ and $7.7 - 20 \ \mu \rm m$. We justify this by the similarity of the infrared absorption spectrum between NH$_4$SH and NH$_4$CN ice (see fig. 1 in \citealt{Slavicinska_2025}). Similarly, for NH$_3$, which condenses at even lower temperatures ($\sim 150$ K), we combine two data sets. We use \cite{Hudson_2022} from $1.67 - 16.67 \mu \rm m$ and \cite{Trotta_1996} from $16.67 - 50 \ \mu \rm m$. For all remaining condensates, the optical constants are taken from the \texttt{LX-MIE} and \texttt{GGChem} repositories.

We emphasize that the optical data for condensates generally come from laboratory measurements at a single temperature. Ideally, we would use optical constants over a range of temperatures, but due to the lack of such experimental data, this study is limited to single-temperature measurements for each condensate. We hope that this will improve in the future.

\subsubsection{Free-free and bound-free absorption}
When the abundance of free electrons becomes non-negligible, the absorption from free-free interactions (inverse Bremsstrahlung) must be taken into account. Table \ref{table:BF_FF_opacity} summarized the free-free interactions considered in this work, along with the relevant references. At long wavelengths ($\lambda > 10 \ \rm \mu m$), we estimate their cross-sections using the $\lambda^2$-scaling as predicted by \cite{Johnston_1967}. 

In addition, we account for the bound-free absorption (photoionization) by the negative hydrogen ion (H$^-$). This process and its reference are also listed in Table \ref{table:BF_FF_opacity}.

\subsubsection{Rayleigh and Thomson scattering}
In addition to absorption cross-sections, we account for scattering cross-sections of several species. We include Rayleigh scattering cross-sections of CO$_2$ (\citealt{Sneep_2005, Thalman_2014}), CO (\citealt{Sneep_2005}), H$_2$ \citep{Cox_2000}, H \citep{Lee_2004}, He (\citealt{Sneep_2005, Thalman_2014}), N$_2$ (\citealp{Sneep_2005, Thalman_2014}), O$_2$ (\citealp{Sneep_2005, Thalman_2014}). We also include Thomson scattering by free electrons \citep{astropy_2022}.

\subsubsection{Opacity spectra}
Before discussing mean opacities, it is instructive to present the monochromatic opacities to understand where they are important. Fig. \ref{fig:opacity_spectra} presents spectra at 1 bar for a solar composition across a range of temperatures, illustrating how different physical processes contribute to the total opacity. 

The black curves represent contributions from molecules and atoms in the gas phase. At $T \lesssim 1000 , \rm K$, the opacity is dominated by H$_2$O, CH$_4$, and NH$_3$, which absorb mainly in the infrared. At higher temperatures, metal hydrides and oxides form, as well as atomic species, and begin to add significant opacity in the optical. In general, molecular and atomic opacities remain important at all temperatures, though their relative influence decreases at higher pressures. 

Continuum sources (scattering, free-free and bound-free absorption, CIA) also play a role. Scattering is most important at short wavelengths ($\lambda \lesssim 1 , \mu$m), particularly at low temperatures. Free–free and bound–free processes become relevant only above $T \gtrsim 1800 \ \rm K$, when free electrons are abundant. CIA appears relatively minor in Fig. \ref{fig:opacity_spectra}, but at higher pressures it will contribute substantially across all temperatures.

Finally, cloud opacities are comparatively flat and featureless but can contribute substantially under certain conditions. Here at  $T = 200 \ \rm K$ (upper left panel), H$_2$O and NH$_4$SH clouds dominate, while at $T = 1800 \ \rm K$ (middle right panel) Fe clouds provide significant opacity. Since condensation is modelled using the rainout approach, condensates do not remain important opacity sources at all altitudes, due to the their gravitational settling which removes them from the overlying atmosphere. As a result, no clouds are present in the $T = 500 \ \rm K$ panel, since condensates such as NaCl and Fe settle at higher temperatures. Moreover, at sufficiently high temperatures, condensation cannot occur.

\subsection{Mean opacities}

To avoid solving the radiative transfer equation for each photon wavelength, stellar, and planetary evolution models typically use mean (or grey) opacities. They correspond to a single number that quantifies how a medium absorbs and scatters radiation over all wavelengths at a pressure-temperature point. The most commonly used mean opacities are the Rosseland mean $\kappa_{\rm R}$ and Planck mean $\kappa_{\rm P}$. They are defined as

\begin{equation}\label{eq:rosseland_opacity}
    \frac{1}{\kappa_{\rm R}} = \frac{\int_0^\infty \kappa_{\lambda}^{-1} \frac{dB_\lambda}{dT} d\lambda}{\int_0^\infty \frac{dB_\lambda}{dT} d\lambda},
\end{equation}

and

\begin{equation}\label{eq:rosseland_opacity}
    \kappa_{\rm P} = \frac{\int_0^\infty \kappa_{\lambda} B_\lambda d\lambda}{\int_0^\infty B_\lambda d\lambda},
\end{equation}

where $\kappa_\lambda$ is the wavelength dependent opacity and $B_\lambda$ is the Planck function. The solid and grey dotted lines in Fig. \ref{fig:opacity_spectra} mark the wavelengths at which $B_\lambda$ and $dB_\lambda/dT$ reach their maxima for the local gas temperature $T_{\rm g}$. This reflects how increasing temperature shifts the weighting of $\kappa_{\rm R}$ and $\kappa_{\rm P}$ toward progressively shorter wavelengths, reducing the relative importance of long-wavelength absorbing species. 

We re-emphasize that in general, $\kappa_{\rm R}$ is used in regions of high optical depth where the diffusion approximation holds. Under these conditions, it can be shown that the radiative temperature gradient depends explicitly on $\kappa_{\rm R}$ \citep{Kippenhahn_1994}. In contrast, $\kappa_{\rm P}$ is used to describe the absorption of optically thin material. It is particularly useful for quantifying energy deposition in a cold gas irradiated by a hotter source, which is a common situation in circumstellar environments. In such cases, it is convenient to weight $\kappa_{\rm P}$ using the radiation temperature of the hotter source $T_{\rm eff}$ rather than the local gas temperature $T_{\rm g}$. This is a quantity that is often used in analytical models for irradiated planetary atmospheres \citep{Guillot_2010}. Hereafter, we refer to mean opacities evaluated at $T_{\rm g}$ as local $\kappa_{\rm R}$, $\kappa_{\rm P}$ and those evaluated at $T_{\rm eff}$ as non-local $\kappa_{\rm R}$, $\kappa_{\rm P}$.


The resolution in $\kappa_\lambda$ required for computing the two mean opacities differs substantially. $\kappa_{\rm R}$ is a harmonic mean, so it is dominated by opacity minima. This means resolving the cores of molecular and atomic absorption lines is not essential and capturing the line wings is sufficient. In practice, this can be achieved with a resolution of $ R = \lambda / \Delta\lambda \approx 10^4$ \citep{Malygin_2014, Siebenaler_2025}.
In contrast, $\kappa_{\rm P}$ is an arithmetic mean dominated by opacity maxima. Hence, resolving the line centres of molecular and atomic absorption features is necessary, which at low pressures requires significantly higher spectral resolution than for $\kappa_{\rm R}$.
In our calculations, $\kappa_{\rm R}$ is computed on a wavenumber grid with spacing $0.1\ \mathrm{cm^{-1}}$ for $\lambda < 10\ \mu\mathrm{m}$ and $0.01\ \mathrm{cm^{-1}}$ for $\lambda \ge 10\ \mu\mathrm{m}$. This ensures a resolution of at least $R = 10^4$ for $\lambda < 100 \ \rm \mu m$. For $\kappa_{\rm P}$, we adopt a spacing of $0.005\ \mathrm{cm^{-1}}$ or $1/4$ of a line width, whichever is larger, which is similar to what is done in F08. However, special care needs to be taken for the Na D and K I resonance doublets, whose Lorentz component can be characterized by a FWHM near $\sim 10^{-7} \ \rm cm^{-1}$ at pressures $\sim 1 \ \rm \mu bar$. Hence, to ensure that these features are resolved, we adopt a wavenumber spacing as low as $\sim 10^{-8} \ \rm cm^{-1}$ around the Lorentz core. Undersampling the Lorentz core of these features leads to significant overestimates of $\kappa_{\rm P}$. At higher pressures ($P > 0.1\ \mathrm{bar}$), we find that the resolution used for $\kappa_{\rm R}$ is also sufficient for computing $\kappa_{\rm P}$.

We computed $\kappa_{\rm R}$ and $\kappa_{\rm P}$ for metallicities $\textrm{[M/H]} = -0.5, -0.3, 0, +0.3, +0.5, +0.7, +1.0, +1.5, +1.7$. Solar abundances are taken to be the present-day solar photospheric values from \cite{Asplund_2021}. In addition we impose a fixed helium to hydrogen mass ratio of $Y/X=0.326$ according to the present-day solar photosphere values. We provide separate tables for cloud-free mean opacities and for cloudy mean opacities. For each table, we calculate the mean opacities at different weighting temperatures, considering both $T_{\rm g}$ and a range of $T_{\rm eff}$. In the case where clouds are considered, we assume the same value of $r_{\rm g}$ for each condensate specie in a given table, as outlined in Section \ref{sec:grain_opacities}.

\section{Results} \label{sec:results}

In this section, we present our mean opacity tables which were calculated using the rainout chemistry approach. Tables \ref{table:cloud_free_table} and \ref{table:cloudy_table} give a description of the content of the tables. All the mean opacity data are available in the Zenodo repository\footnote{\url{https://doi.org/10.5281/zenodo.17418093}}. 

We begin by introducing our cloud-free mean opacities, which is followed by the mean opacities accounting for clouds.

\subsection{Cloud-free mean opacities}

\subsubsection{Rosseland mean}

\begin{figure}
	\includegraphics[width=1\columnwidth]{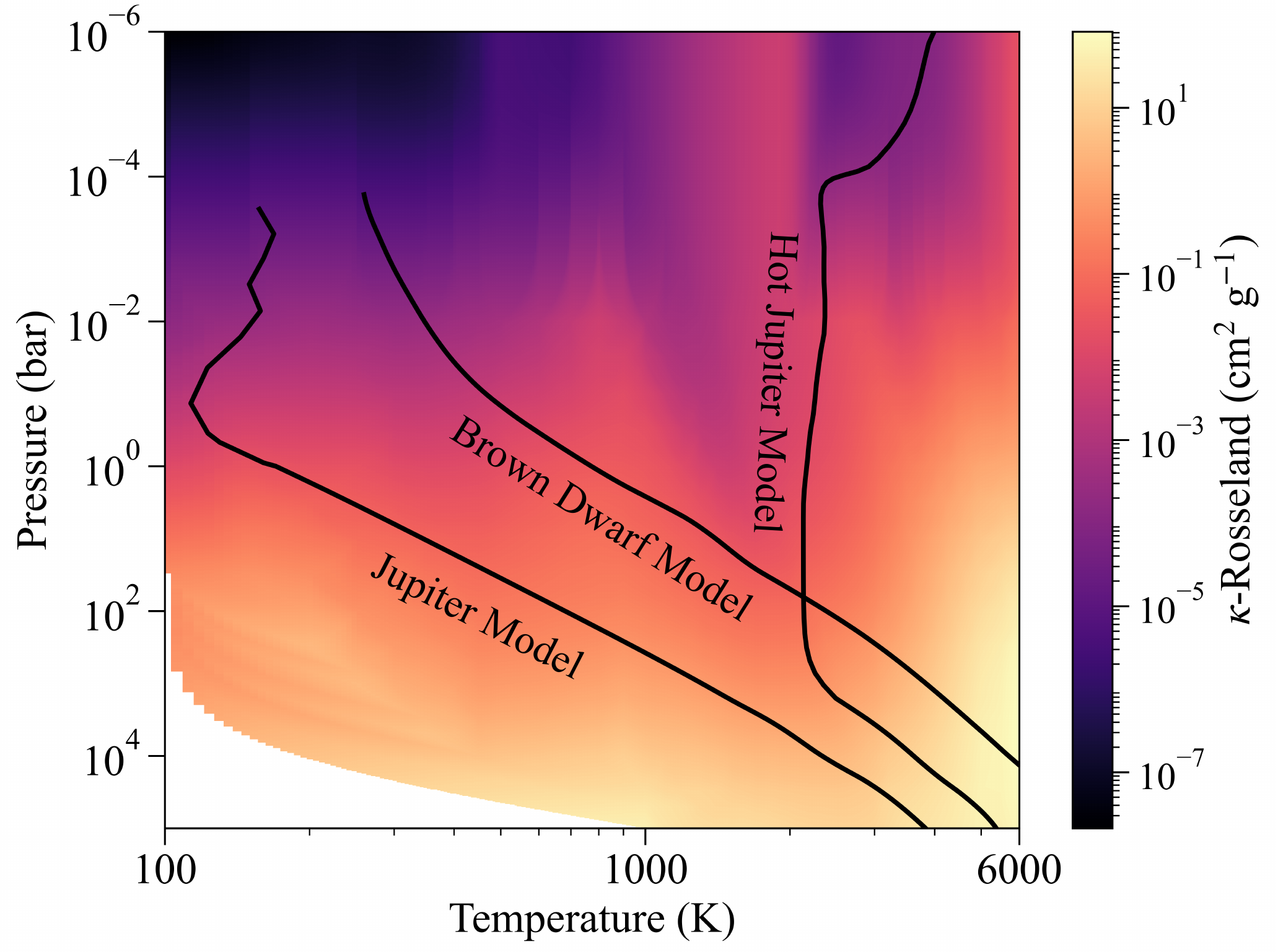}
    \caption{Local Rosseland mean opacity $\kappa_{\rm R}$ computed for a solar composition over the grid of temperature and pressure considered in this work. In black we show example thermal profiles which fall within our opacity grid. The Jupiter model is taken from \citet{Siebenaler_2025}, the brown dwarf model is from \citet{Marley_2021}, and the hot Jupiter model is from \citet{Goyal_2020}.}
   \label{fig:opacity_map}
\end{figure}

\begin{figure*}
	\includegraphics[width=0.9\textwidth]{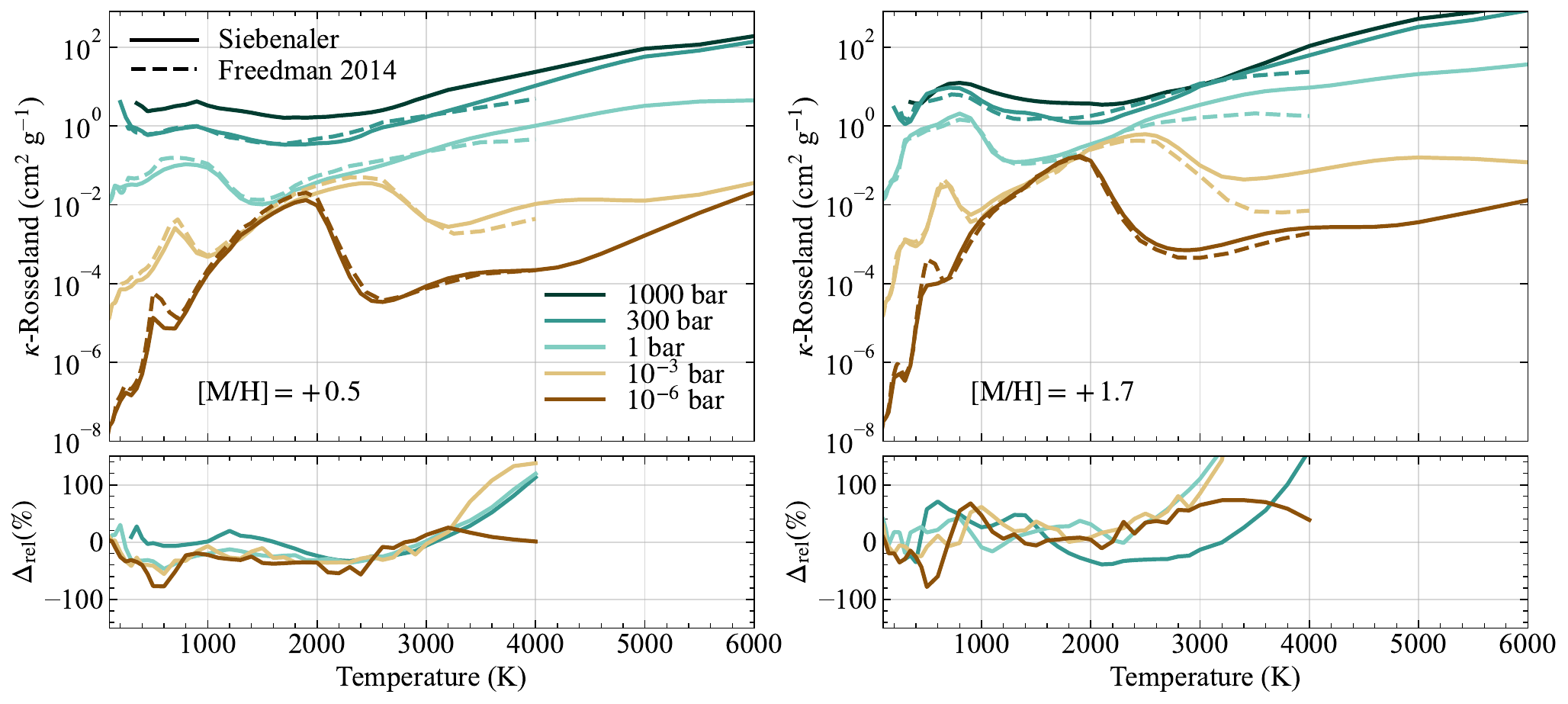}
    \caption{Upper panels: Local Rosseland mean opacity $\kappa_{\rm R}$ as a function of temperature for various fixed pressures. The solid curves correspond to the data from this study, while the dashed curves come from \citet{Freedman_2014}. Lower panels: The relative difference in $\kappa_{\rm R}$ between our data and that of \citet{Freedman_2014} at fixed pressures. Right panels: Apply to a metallicity of $[\rm M/H] = +0.5$. Left panels: Apply to a metallicity of $[\rm M/H] = +1.7$.}
   \label{fig:ross_comp_freedman_+0.5_+1.7}
\end{figure*}

Fig. \ref{fig:opacity_map} shows our local cloud-free $\kappa_{\rm R}$ map for a solar composition. We also include model pressure–temperature profiles for a variety of substellar objects, demonstrating the applicability of our tables across different types of atmospheres and interiors. As expected, the gaseous opacity generally increases with temperature and pressure. This trend arises from several factors. At higher temperatures, more atomic and molecular absorption transition are available, increasing their monochromatic opacity $\kappa_{\rm \lambda}$, while the rising abundance of free electrons enhances free–free and bound–free opacities. At higher pressures, pressure broadening redistributes absorption from line cores into the wings, increasing $\kappa_{\rm R}$, and CIA opacities also becomes more important.

The upper left panel in Fig. \ref{fig:ross_comp_freedman_+0.5_+1.7} shows our local cloud-free $\kappa_{\rm R}$ (solid curves) at a metallicity of  $\textrm{[M/H]} = +0.5$, compared with F14. The relative difference between the two data sets is given in the lower left panel. In general, differences are around $\sim 40\%$, with our $\kappa_{\rm R}$ being smaller than F14. At high temperatures ($\gtrsim 3000 \ \rm K$) the discrepancies grow and can exceed $100\%$, with our $\kappa_{\rm R}$ being larger than F14. This is primarily caused by metal hydrides (CaH, CrH, FeH, MgH, NaH, SiH, and TiH), oxides (CaO, MgO, SiO, TiO, and VO), hydroxides (CaOH), and atomic species (Na, K, Fe, Ni, Cr, Li, Ca, and Mg)  that were included in this study and dominate at short wavelenghts ($\lambda \lesssim 1 \ \rm \mu m$). In contrast, F14 includes far fewer short-wavelength absorbing species. The origin of the smaller differences at lower temperatures is challenging to trace without access to the original cross-section and chemistry data used in F14. Potential causes include differences in line lists, pressure-broadening treatments, line wing cut-offs, or CIA data. 

The differences in $\kappa_{\rm R}$ between the two data sets increases with metallicity. At $\textrm{[M/H]} = +1.7$, relative differences can exceed $500\%$ at $T \gtrsim 3000 \ \rm K$, as shown in the right panel of Fig. \ref{fig:ross_comp_freedman_+0.5_+1.7}.

Finally, we also compared our low-density $\kappa_{\rm R}$ values with those of \cite{Ferguson_2005} (see Appendix \ref{sec:appendix_equ_cond}). The agreement is good which further supports the robustness of our calculations.

\subsubsection{Planck mean}

\begin{figure}
	\includegraphics[width=\columnwidth]{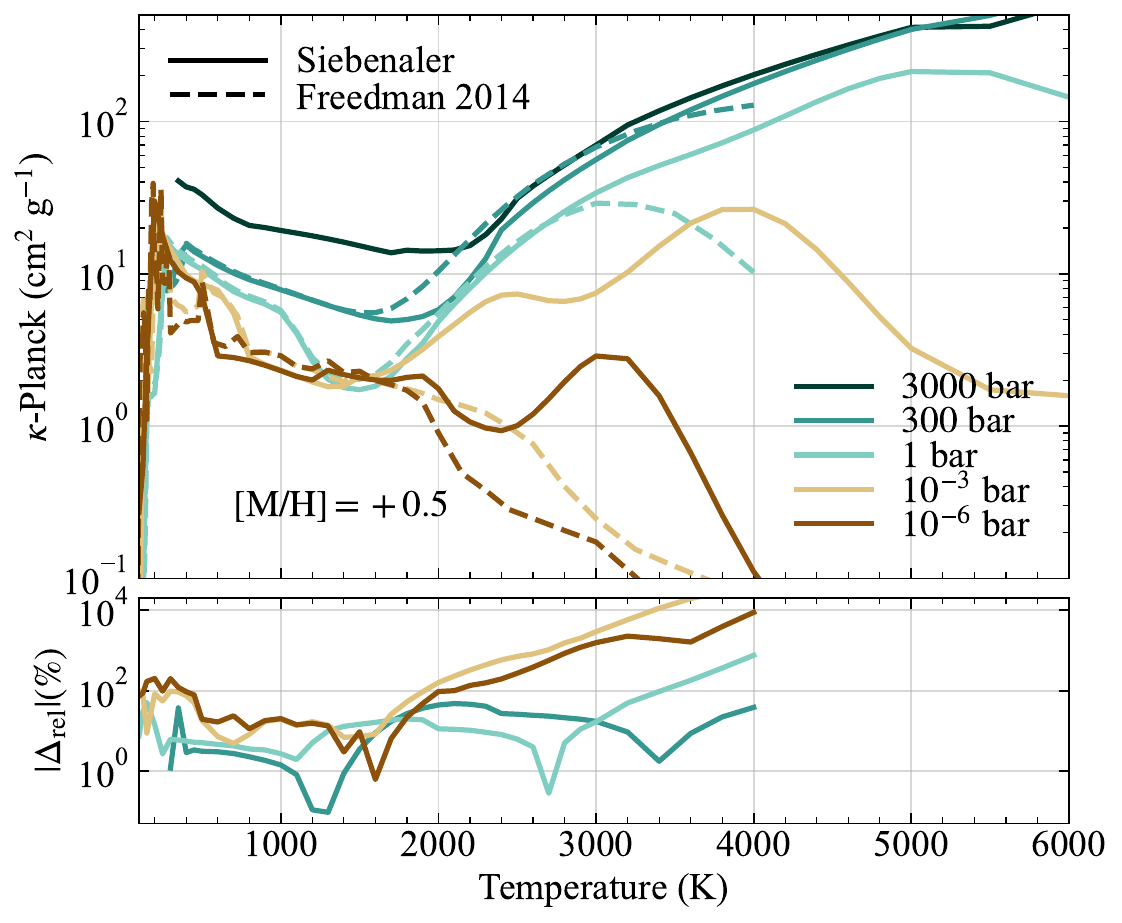}
    \caption{Upper panel: Local Planck-mean opacity $\kappa_{\rm P}$ as a function of temperature for various fixed pressures at a metallicity of $[\rm M/H] = +0.5$. The solid curves correspond to the data from this study, while the dashed curves come from \citet{Freedman_2014}. Lower panel: The relative difference in $\kappa_{\rm P}$ between our data and that of \citet{Freedman_2014} at fixed pressures.}
   \label{fig:planck_comp_freedman_+0.5}
\end{figure}

We compare our local cloud-free $\kappa_{\rm P}$ with F14 at a metallicity of $\textrm{[M/H]} = +0.5$ in Fig. \ref{fig:planck_comp_freedman_+0.5}. In general, the discrepancies between the two data sets are much larger across all temperatures than for $\kappa_{\rm R}$. This is not surprising since $\kappa_{\rm P}$ is weighted toward opacity maxima, making it more sensitive to additional opacity sources and updated line lists than $\kappa_{\rm R}$. At higher pressures ($P \gtrsim 0.3 \ \rm bar$), these differences decrease, although they remain substantial at high temperatures. 

\begin{figure}
	\includegraphics[width=\columnwidth]{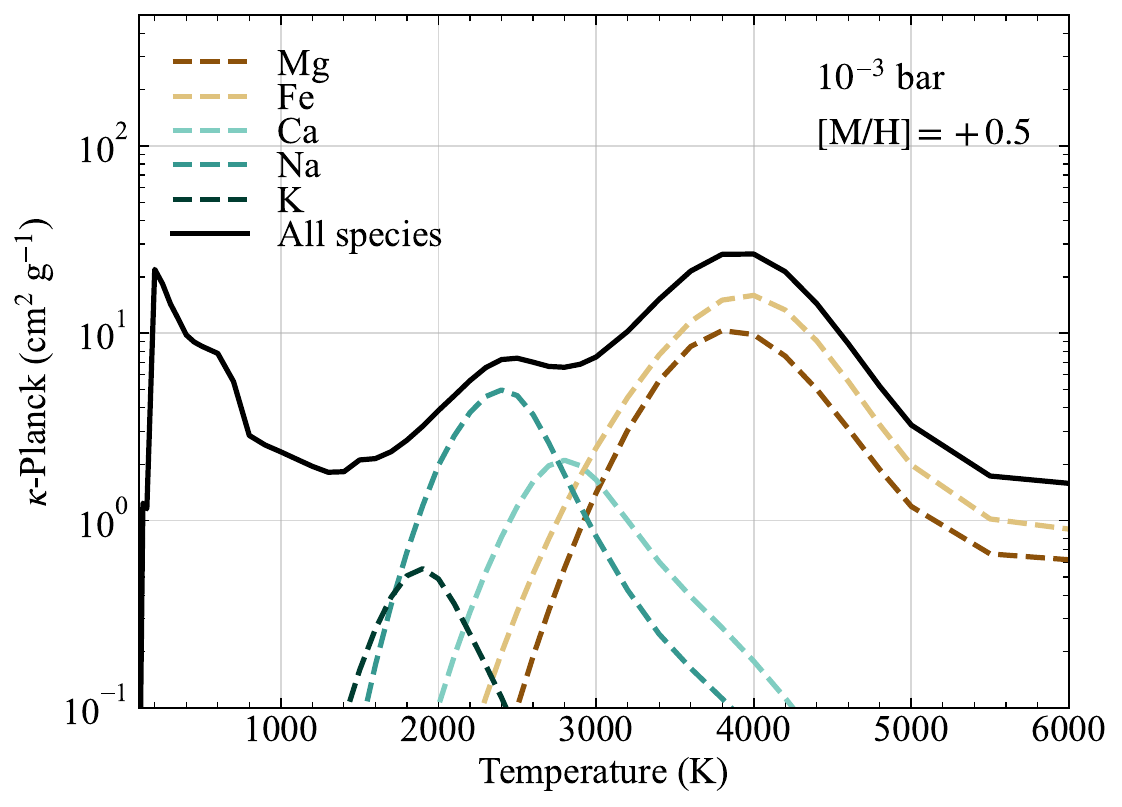}
    \caption{Local $\kappa_{\rm P}$ as a function of temperature at a fixed pressure of $10^{-3} \ \rm bar$ and a metallicity of $\rm [\rm M/H] = +0.5$. The black solid curve gives the total $\kappa_{\rm P}$ from all the species considered in this study. The dashed colored curves corresponds to the contributions of individual species to the total $\kappa_{\rm P}$.}
   \label{fig:planck_species_+0.5}
\end{figure}

The largest relative differences occur at low pressures and high temperatures, exceeding $1000\%$ at $T \gtrsim 2000 \ \rm K$, and are primarily caused by atomic species. Fig. \ref{fig:planck_species_+0.5} shows the contributions of several atomic species to the local $\kappa_{\rm P}$ at $10^{-3} \ \rm bar$, which is where the largest deviations from F14 are observed. Na and K show significant contributions to $\kappa_{\rm P}$ around $2000 \ \rm K$, which is caused by the extremely strong peaks of the Lorentz core of the Na D and K I resonance lines as predicted by the theory of \cite{Allard_2016, Allard_2019}. In contrast, F14 used the \cite{Burrows_2000} approach for the opacities of Na and K, which appears to produce substantially different $\kappa_{\rm P}$ near $2000 \ \rm K$ at low pressures. At even higher temperatures ($T \gtrsim 3000 \ \rm K$), other atomic species such as Ca, Mg, and Fe contribute significantly to $\kappa_{\rm P}$. Although Ca and Mg have relatively few spectral lines (22339 lines for Ca; 835 lines for Mg), they exhibit resonance absorption lines that are extremely strong and therefor dominate $\kappa_{\rm P}$. The most important are the Ca I line at $0.4226 \ \rm \mu m$ and Mg I line at $0.2852 \ \rm \mu m$. Ab initio calculations on their lineshapes \citep{Allard_2018, Blouin_2019} show that their linewings extend far beyond the $R_{\rm cut}$ value adopted in this study. Nevertheless, we tested different $R_{\rm cut}$ values for these resonance lines and find that it has no noticeable impact on $\kappa_{\rm P}$ \footnote{We also tested the impact of different $R_{\rm cut}$ values of the Ca I and Mg I resonance line on $\kappa_{\rm R}$ and find that it is negligible.}. This is because $\kappa_{\rm P}$ mostly depends on the integrated line strength, which we ensure is conserved when changing $R_{\rm cut}$. On the other hand, Fe is characterized by more absorption lines (126288 lines) and has no single absorption line that dominates $\kappa_{\rm P}$. 

\begin{figure}
	\includegraphics[width=\columnwidth]{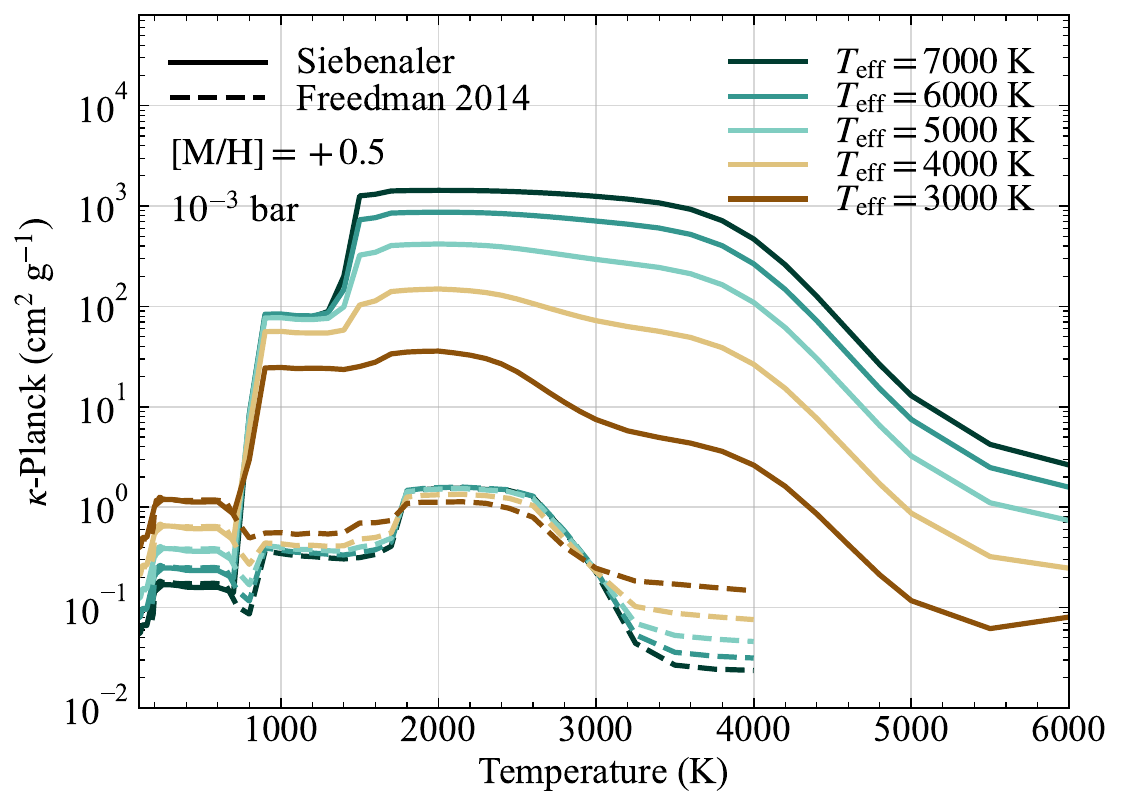}
    \caption{Non-local $\kappa_{\rm P}$ as a function of temperature at a fixed pressure of $10^{-3} \ \rm bar$ and a metallicity of $\rm [\rm M/H] = +0.5$. The solid curves correspond to the data from this study, while the dashed curves come from \citet{Freedman_2014}. Each curve corresponds to a different weighting temperature $T_{\rm eff}$.}
   \label{fig:planck_Teff}
\end{figure}

Given the differences in the local $\kappa_{\rm p}$ between our data set and F14, there are also significant differences in the non-local $\kappa_{\rm p}$ (evaluated at $T_{\rm eff}$). Fig. \ref{fig:planck_Teff} compares our non-local $\kappa_{\rm p}$ to F14 at a pressure of $10^{-3}$ bar. The agreement is very good for $T \lesssim 900 \ \rm K$, however at higher $T$ the increasing abundance of Na and K leads to substantial deviations. The jump in non-local $\kappa_{\rm p}$ near $T \sim 1300 \ \rm K$ is primarily caused by Mg and Fe becoming abundant.

Lastly, similar to $\kappa_{\rm R}$, the discrepancies in $\kappa_{\rm P}$ between our data set and that of F14 increase at higher metallicity.

\begin{figure*}
	\includegraphics[width=0.9\textwidth]{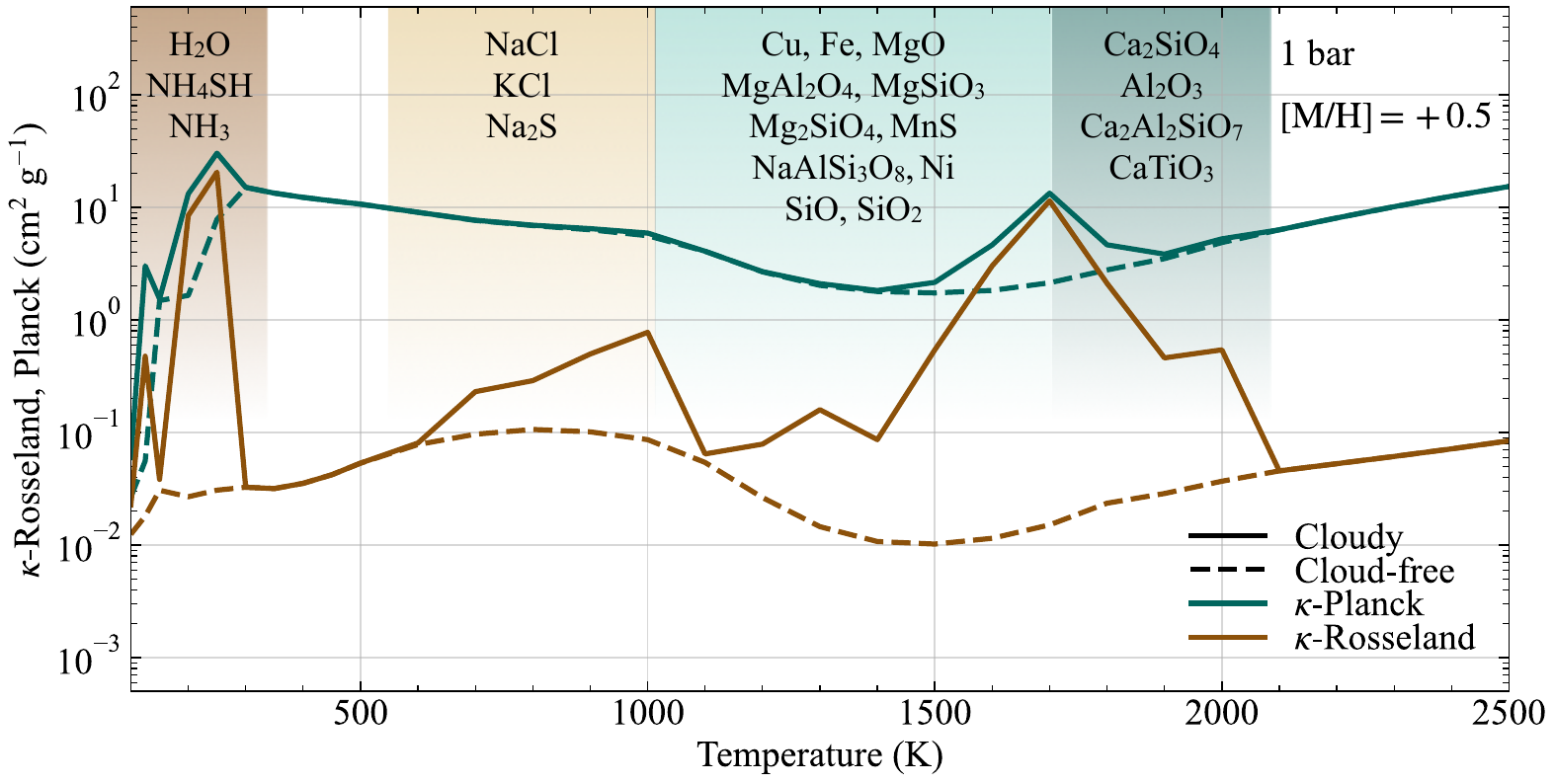}
    \caption{Local $\kappa_{\rm R}$ and $\kappa_{\rm P}$ as function of temperature at $1 \rm \ bar$ and a metallicity $\rm [M/H] = +0.5$. The solid curves correspond to the cloudy mean opacities with a mean particle radius $r_{\rm g} = 1\rm \ \mu m$, while the dashed curves are the cloud-free mean opacities. The shaded regions indicate which condensates exist at different temperatures. 
    }
   \label{fig:ross_planck_clouds_1micron}
\end{figure*}

\begin{figure}
\includegraphics[width=\columnwidth]{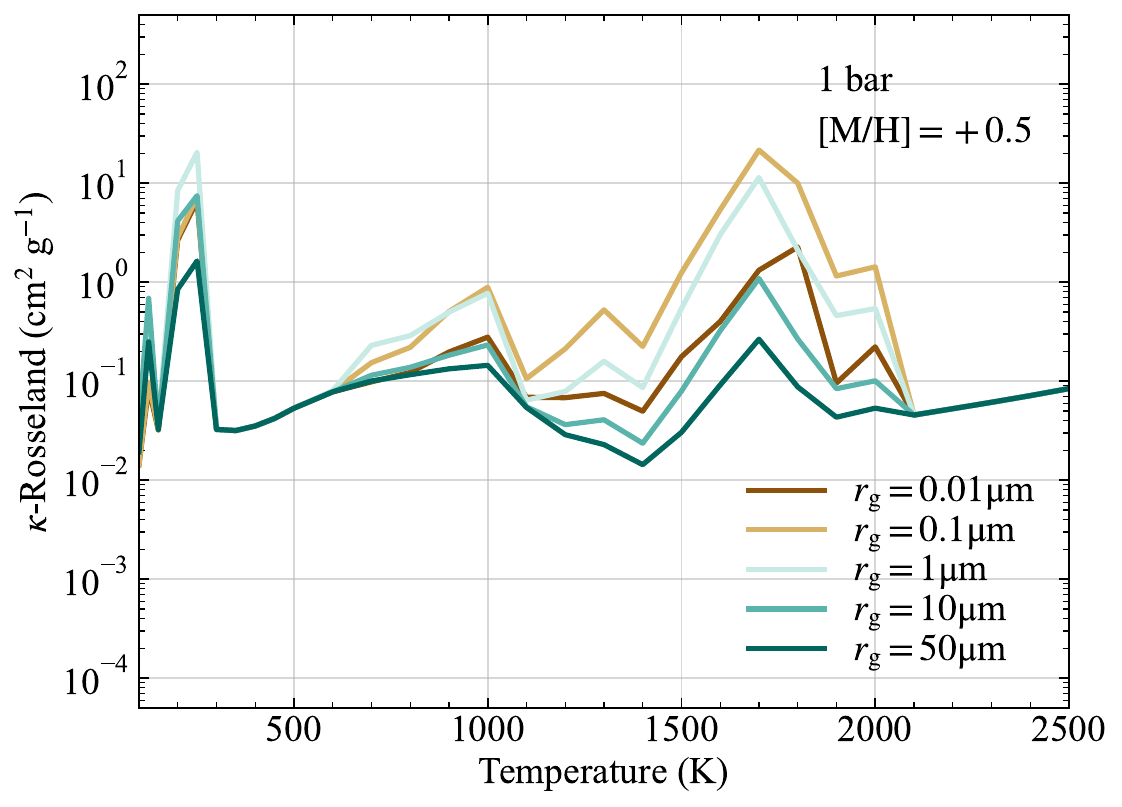}
    \caption{Local cloudy $\kappa_{\rm R}$ as function of temperature at $10^{-3} \rm \ bar$ and a metallicity $\rm [M/H] = +0.5$. Each curve assumes a different value for the mean cloud particle radius $r_{\rm g}$. }
   \label{fig:ross_clouds_radii}
\end{figure}

\subsection{Cloudy mean opacities}

We now present the cloudy mean opacities. Fig. \ref{fig:ross_planck_clouds_1micron} shows local $\kappa_{\rm R}$ and $\kappa_{\rm P}$ at a pressure of $1 \ \rm bar$ and $\textrm{[M/H]} = +0.5$, assuming a mean particle radius of $r_{\rm g} = 1 \ \mu \rm m$ (solid curves). For comparison, the cloud-free mean opacities are shown as dashed curves. At $T \lesssim 2000 \ \rm K$, clouds strongly affect $\kappa_{\rm R}$, while their effect on $\kappa_{\rm P}$ is less pronounced. As illustrated in Fig. \ref{fig:opacity_spectra}, cloud opacities are relatively flat across wavelength, which fills in opacity minima and substantially increases $\kappa_{\rm R}$, while leaving opacity maxima, and thus $\kappa_{\rm P}$ largely unchanged, unless the cloud abundance is very high. Unlike in \cite{Ferguson_2005} and \cite{Marigo_2024}, our cloudy mean opacities do not show a pronounced plateau in $\kappa_{\rm R}$ extending to $\sim 2000 \ \rm K$. This difference arises because we adopt the rainout approach to model condensation, where condensates settle into distinct layers rather than remaining mixed throughout the entire atmosphere. 

The shaded regions in Fig. \ref{fig:ross_planck_clouds_1micron} indicate where condensates contribute to the mean opacity. These can be broadly grouped into four classes: low-$T$ (water and N-bearing condensates), intermediate-low-$T$ (mainly salts), intermediate-high-$T$ (mainly Mg- and Si-bearing condensates), and high-$T$ (mainly Ca- and Al-bearing species). Since condensation curves are pressure dependent, the exact location of each group and condensate shifts with pressure, generally moving to higher temperatures as pressure increases. At $T \gtrsim 2800 \ \rm K$, no condensates form regardless of pressure, and the mean opacities are determined solely by gaseous species.

The value of $r_{\rm g}$ strongly affects the extinction efficiency of cloud particles. In Fig. \ref{fig:ross_clouds_radii}, we show local $\kappa_{\rm R}$ for different $r_{\rm g}$ values. Changing $r_{\rm g}$ can modify $\kappa_{\rm R}$ by up to two orders of magnitude. We note that the effect on $\kappa_{\rm P}$ remains much smaller. Overall, we find that cloud opacities are maximized for particle radii around $r_{\rm g} \sim 0.1 - 1 \ \rm \mu m$.

\section{Discussion} \label{sec:discussion}

\subsection{Impact on planetary evolution}

\begin{figure*}
	\includegraphics[width=1\textwidth]{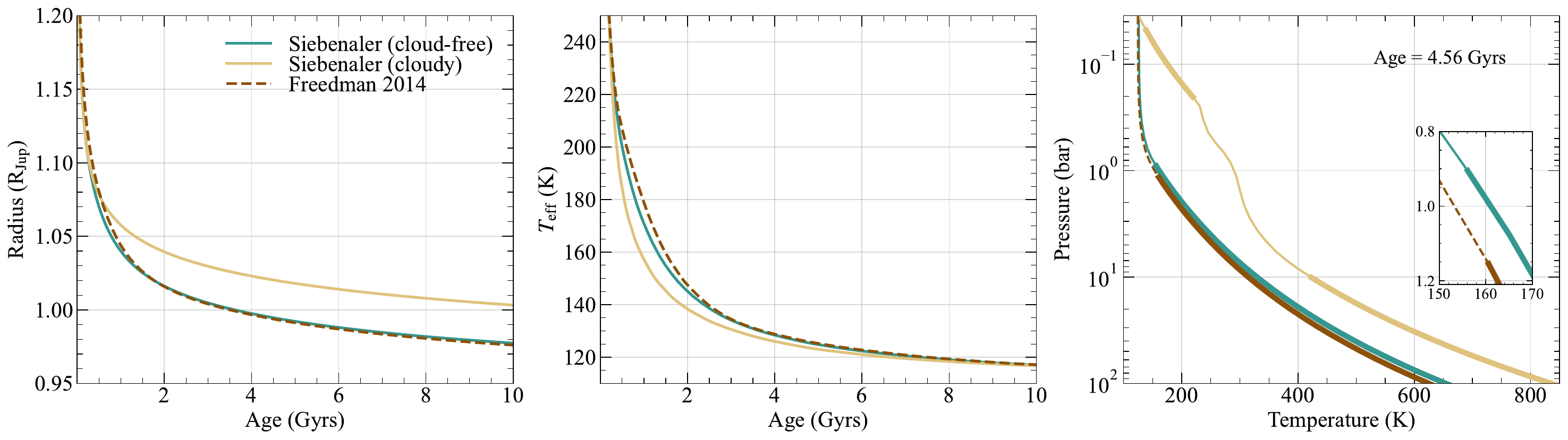}
    \caption{Evolutionary calculations of a Jupiter-like planet for different opacity tables at a metallicity of $\rm [M/H] = +0.5$. Solid cyan curves correspond to the cloud-free table computed in this work, solid yellow curves apply to the cloudy table with $r_{\rm g} = 1 \ \rm \mu m$, and dashed brown curves use the \citet{Freedman_2014} table. Left panel: Planetary radius (units in Jupiter radius $\rm R_{Jup}$) as a function of age. Middle panel: Planetary effective temperature $T_{\rm eff}$ with age. Right panel: Thermal profiles at an age of 4.56 Gyrs. The thicker regions indicate convective layers. The inset corresponds to a zoom-in around the 1-bar level.}
   \label{fig:jup_ev_+1.5}
\end{figure*}

In this work, we have computed new mean opacity tables tailored for giant planet modeling. Here, we assess how these tables can affect planetary evolution by running interior models of a Jupiter-like planet with \texttt{CEPAM} \citep{Guillot_1995}. The planet is assumed to have a $10 \ \rm M_\oplus$ core composed of $50\%$ rock and $50\%$ ice, surrounded by a homogeneous H–He envelope of protosolar composition. For simplicity, the atmospheric boundary is modelled with the Eddington approximation \citep{Eddington_1926}. 

The left and middle panels of Fig. \ref{fig:jup_ev_+1.5} show the evolution of the planetary radius and effective temperature $T_{\rm eff}$ at $\textrm{[M/H]}=+0.5$. Using our cloud-free opacity table (solid dark cyan) or F14 (dashed brown) produces nearly identical evolutionary tracks. This agreement is expected, since differences between the two data sets at $\textrm{[M/H]}=+0.5$ become significant only at high temperatures (see Fig. \ref{fig:ross_comp_freedman_+0.5_+1.7}). Such temperatures are not reached at low pressures and above the radiative-convective boundary during the evolution of our Jupiter-like planet, which is the region that will control the planet's cooling rate. The right panel of Fig. \ref{fig:jup_ev_+1.5} compares thermal profiles at an age of 4.56 Gyrs. The thick curves indicate the regions where the planet is convective. Although the radius and $T_{\rm eff}$ evolutions are very similar, the interior of the planet cools slightly less efficiently with our opacities, raising the temperature near 1 bar by $\sim 10 \ \rm K$ and shifting the radiative–convective boundary to lower pressures.

The effect of cloud opacities is shown by the solid yellow curve, assuming $r_{\rm g} = 1 \ \rm \mu m$. For the first $\sim 0.4 \ \rm Gyr$, the evolution follows the cloud-free track. Once the atmosphere cools enough for H$_2$O and NH$_4$SH to condense, the opacity at low pressures rises sharply. Throughout the remainder of the evolution, H$_2$O and NH$_4$SH clouds persist, slowing the planet’s cooling and increasing its Kelvin–Helmholtz time-scale. Around $\sim 2.5 \ \rm Gyrs$, the planet will have cooled enough to also form NH$_3$ clouds. By the end of the evolution, the radius is inflated by $\sim 3\%$ relative to the cloud-free case. During the early contraction, the planet also appears significantly fainter, with $T_{\rm eff}$ lower by about $15 \ \rm K$ at 1 Gyr. The right panel further shows how clouds modify the thermal structure and act as heat traps. The interior temperature is higher by $\sim 200 \ \rm K$ at 100 bar, and a convective layer forms in the atmosphere at the location of the cloud deck ($\sim 0.05 - 0.2 \ \rm bar$).

These results demonstrate that our opacity tables can influence planetary evolution models, particularly when cloud opacities are included. The role of clouds as heat traps also has important implications on the inferred bulk metallicity of the planet. To reproduce an observed radius, interior models that include clouds must assume a larger total heavy-element mass than cloud-free models in order to compensate for the slower contraction. We note that the impact of our cloud-free tables is expected to be more pronounced for hotter planets or those with higher metallicity.

We emphasize that real Jupiter is significantly more complex. Observations (e.g. \citealt{Guillot_2020, Biagiotti_2025}) show that its cloud structure is far more intricate than predicted by equilibrium chemistry alone. For exoplanets, however, cloud properties are poorly constrained, and our tables provide a useful first-order approach to incorporate their effect into evolutionary models. We also note that the thermal profiles here are based on radiative gradients computed with Rosseland-mean opacities, which is valid in optically thick regions where the diffusion approximation holds. This assumption may break down in the upper layers of the interior model used here.

\subsection{High-pressure opacities}

Mean opacities across a wide pressure range ($10^{-6}$–$10^5$ bar) have been computed in this study. Non-ideal effects become increasingly important at high pressures, and in this section we want to acknowledge the uncertainties associated with our high-pressure opacities.  
 
\begin{figure}
	\includegraphics[width=0.9\columnwidth]{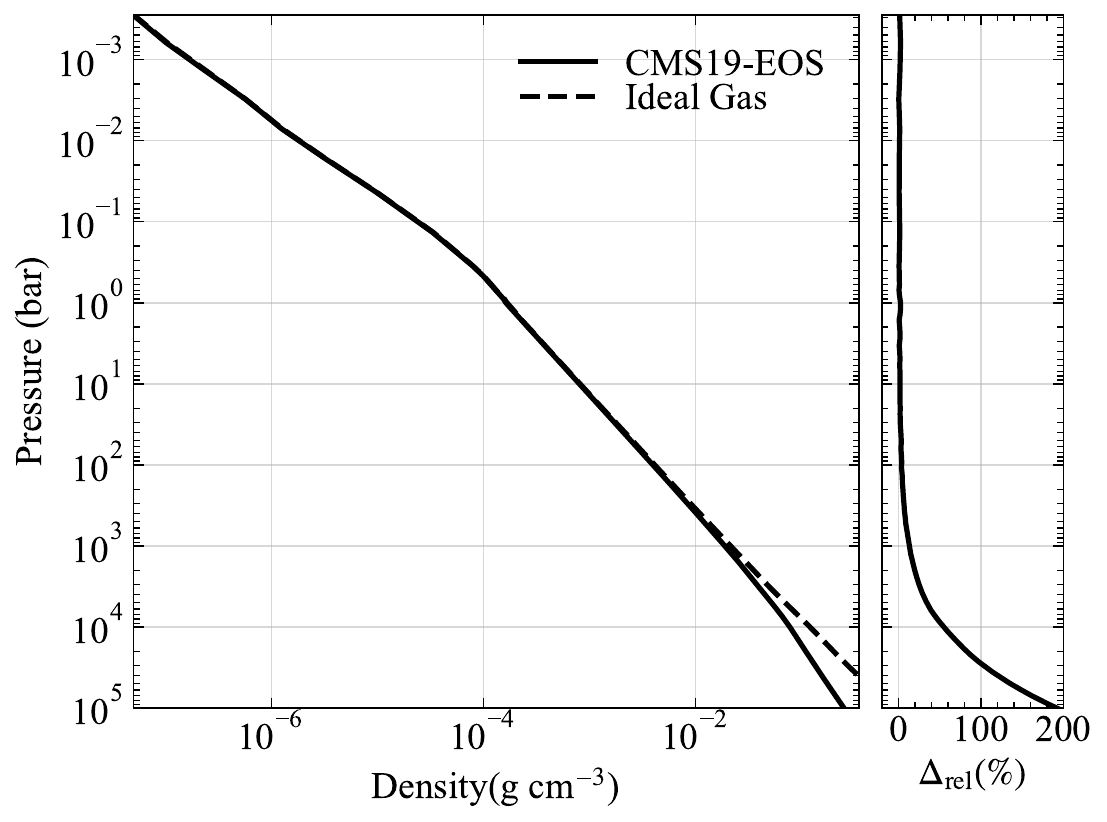}
    \caption{Left panel: Mass density as a function of pressure along a Jupiter thermal profile taken from \citet{Siebenaler_2025}. The dashed curve corresponds to the density from the ideal gas law computed using \texttt{GGChem} and assuming a solar composition. The solid curve corresponds to the density computed using the CMS19 equation of state \citet{Chabrier_2019} assuming a solar composition H-He mixture. Right panel: Relative difference between the density profiles.}
   \label{fig:eos_vs_ideal}
\end{figure}

Our equilibrium chemistry calculations rely on the law of mass action for the formulation of the thermochemical equilibrium, which is derived assuming an ideal gas mixture. At sufficiently high pressures, this assumption will break down, and species-specific equations of state, derived from experiments or ab initio calculations, are required for an accurate thermodynamic description. In Fig. \ref{fig:eos_vs_ideal}, we compare the mass density of a solar H–He mixture along a Jupiter thermal profile using the CMS19 equation of state \citep{Chabrier_2019} with that of an ideal gas, as used in \texttt{GGChem}. Above $\sim 1000 \ \rm bar$, the H–He mixture exhibits non-negligible deviations from ideal-gas behaviour. This suggests that our chemistry calculation could become unreliable around these pressures, given the hydrogen-dominated atmospheres considered here. The impact of including non-ideal equations of state in thermochemical equilibrium calculations is difficult to assess, as equations of state for many species are not available. In addition, high pressures can lead to ionization potential depression \citep{Ecker_1963, Stewart_1966}, which alters the electron abundance. However, based on the analysis of \citet{Marigo_2024}, we expect this effect to be negligible within the parameter space relevant for our mean opacities.

High-pressures will also modify CIA cross-sections. In this study, we only account for two-body collisions, but at sufficiently high pressures, three- or even four-body collisions should be considered. For H$_2$–He collisions, \citet{Dossou_1986} finds that three-body collisions can already become important at densities $\gtrsim 0.04 \ \rm g \ cm^{-3}$, which we can roughly translate to an ideal gas pressure of $\gtrsim 1000 \ \rm bar$. Including higher order CIA terms will increase $\kappa_{\rm R}$ and $\kappa_{\rm P}$. As a result, the CIA opacities used in this study should be regarded as a lower limit at high pressures.

The description of the line profiles can also become problematic at high pressure. At low pressures, the assumption of a Voigt profile to model molecular and atomic absorption lines is adequate. At higher pressures, however, distortions away from the Voigt profile occur due to line mixing, a collisional process that couples different transitions \citep{Levy_1992, Pieroni_2001, Hartmann_2018}. Physically, line mixing allows an absorption line to be produced through an alternative path that involves a collisional transition. While line positions remain unaffected, the population levels and line shape parameters are modified. Line mixing can already become noticeable near $10 \ \rm bar$ for molecules such as CO$_2$, CO, and H$_2$O, as shown by \citet{Ren_2023}. At present, however, its impact on mean opacities cannot be assessed, since data on deviations from Voigt profiles remain scarce and are typically measured for N$_2$ and O$_2$ as broadening agents, appropriate for Earth-like conditions. We note that F08 tested line mixing for H$_2$O and concluded that it does not substantially affect mean opacities, however this requires further assessment. 

The Na D and K I resonance lines pose an additional challenge. At perturber densities exceeding $10^{21} \ \rm cm^{-3}$, we adopt a Voigt profile with an extended wing cut-off, where the HWHM is computed using the impact approximation. However, this approach can underestimate the true broadening at very high densities and thus the contribution of these lines to the mean opacity calculation. Recent theoretical calculations from \cite{Allard_2025} show that the impact theory breaks down at perturber densities $\gtrsim 4 \cdot 10^{21} \ \rm cm^{-3}$ and can underestimate the HWHM by up to a factor of five at densities $\sim 3 \cdot 10^{22} \ \rm cm^{-3}$. This is a manifestation of satellite components \citep{Kielkopf_1979, Kielkopf_1983} becoming increasingly more important at higher perturber densities, shifting the position of the whole profile and introducing non-Lorentzian features. Including such descriptions in our cross-section calculations is out of the scope of this study, and will be deferred to a future study. Based on the results in \cite{Allard_2025}, we expect that the Na D and K I contributions to $\kappa_{\mathrm{R}}$ are likely underestimated in our high-pressure calculations ($\gtrsim 1000 \ \rm bar$).

Given the outlined limitations, our mean opacities should be regarded as most reliable within hydrogen-dominated atmospheres at pressures up to $\sim 1000 \ \rm bar$. Beyond these pressures, deviations from ideal-gas behaviour and higher-order collisional effects may introduce significant, but as of yet unquantified, errors. This highlights the need for improved experimental and theoretical data on equations of state to reliably extend mean opacity calculations to the high-pressure regime.

\section{Conclusions} \label{sec:conclusion}
We have computed $\kappa_{\rm R}$ and $\kappa_{\rm P}$ tables for nine elemental compositions relevant to giant planets. The tables span pressures from 
$10^{-6} - 10^{5} \ \rm bar$ and temperatures from $100 - 6000 \ \rm K$, covering a substantially wider $P-T$ range than other commonly used opacity tables in planetary science. This broad coverage makes them applicable to both the atmospheres and shallow interiors of cold and hot giant planets. Our calculations employ the latest molecular line lists and pressure-broadening parameters for a large number of molecular and atomic species, as well as state-of-the-art treatments of the Na D  and K I resonance lines. In addition, we provide opacity tables that include clouds across a wide range of particle sizes, representing the first publicly available mean opacity tables with cloud contributions tailored to study giant planets.

We benchmarked our cloud-free tables against the widely used F14 data set and find significant deviations in $\kappa_{\rm R}$ for $T \gtrsim 3000 \ \rm K$. This can affect the modeling of hot Jupiters, particularly when interpreting their radiative-convective boundary. These differences can be attributed to the inclusion of a large set of short-wavelength absorbing species in our calculations. Deviations in $\kappa_{\rm P}$ are even more pronounced at high temperatures. This appears to be driven by the inclusion of key atomic species Mg, Fe, and Ca, and updated Na D and K I resonance line profiles. Our new $\kappa_{\rm P}$ values may influence analytical models of the thermal structure of giant planet atmospheres.

We also find that clouds substantially increase $\kappa_{\rm R}$ for $T \lesssim 2800 \ \rm K$, while their effect on $\kappa_{\rm P}$ is weaker. In an example evolution model of a Jupiter-like planet, including clouds produces a $\sim 3\%$ larger radius and a significantly hotter interior. These tables will also be useful to further improve on our understanding of cloudy atmospheres of warm giant planets and how they impact their evolution and thermal structure.

Overall, our tables offer several key improvements over the F14 data set. They incorporate a substantially larger set of absorption sources, use the latest line lists and pressure-broadening data, and fill the long-standing gap in cloudy mean opacity data for giant planets. They can be used with confidence up to $\sim 1000 \ \rm bar$, while higher-pressure values should be treated with caution until improved experimental and ab initio constraints become available. We anticipate that these tables will serve as a valuable resource for studies of planetary atmospheres, interiors, and evolution.


\section*{Acknowledgements}
This project has received funding from the European Research Council (ERC) under the European Union’s Horizon 2020 research and innovation programme (grant agreement no. 101088557, N-GINE). This publication is part of the project ENW.GO.001.001 of the research programme “Use of space infrastructure for Earth observation and planetary research (GO), 2022-1” which is (partly) financed by the Dutch Research Council (NWO). This work used the Dutch national e-infrastructure with the support of the SURF Cooperative using grant no. EINF-11325. We thank Nicole Allard for providing us with the opacity tables of the K I resonance lines perturbed by H$_2$ before they were made publicly available. We also thank Tristan Guillot for insightful discussions.

\section*{Data Availability}

The mean opacity tables are available at \url{https://doi.org/10.5281/zenodo.17418093}.



\bibliographystyle{mnras}
\bibliography{bibliography} 



\appendix

\section{Equilibrium condensation opacities} \label{sec:appendix_equ_cond}

In addition to the mean opacity tables based on the rainout chemistry approach, we have also constructed tables assuming equilibrium condensation. These are more appropriate for modeling low-gravity environments such as protoplanetary discs. The equilibrium-condensation tables cover the same metallicities and cloud particle size distributions as the rainout set.

\begin{figure}
	\includegraphics[width=0.9\columnwidth]{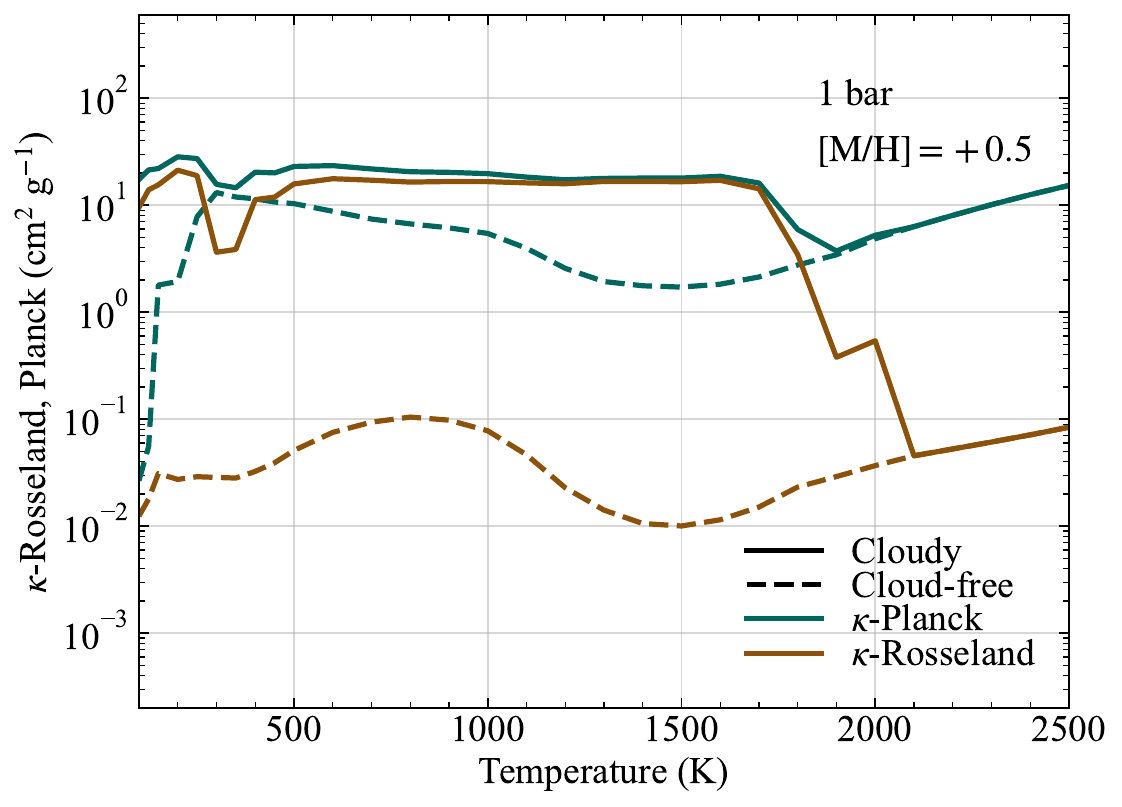}
    \caption{Local $\kappa_{\rm R}$ and $\kappa_{\rm P}$ as function of temperature at $1 \rm \ bar$ and a metallicity $\rm [M/H] = +0.5$. Equilibrium condensation is assumed for the chemistry calculation. The solid curves correspond to the cloudy mean opacities with $r_{\rm g} = 1 \ \rm \mu m$, while the dashed curves are the cloud-free mean opacities.}
   \label{fig:equ_cond_Z+0.5}
\end{figure}

\begin{figure}
	\includegraphics[width=0.9\columnwidth]{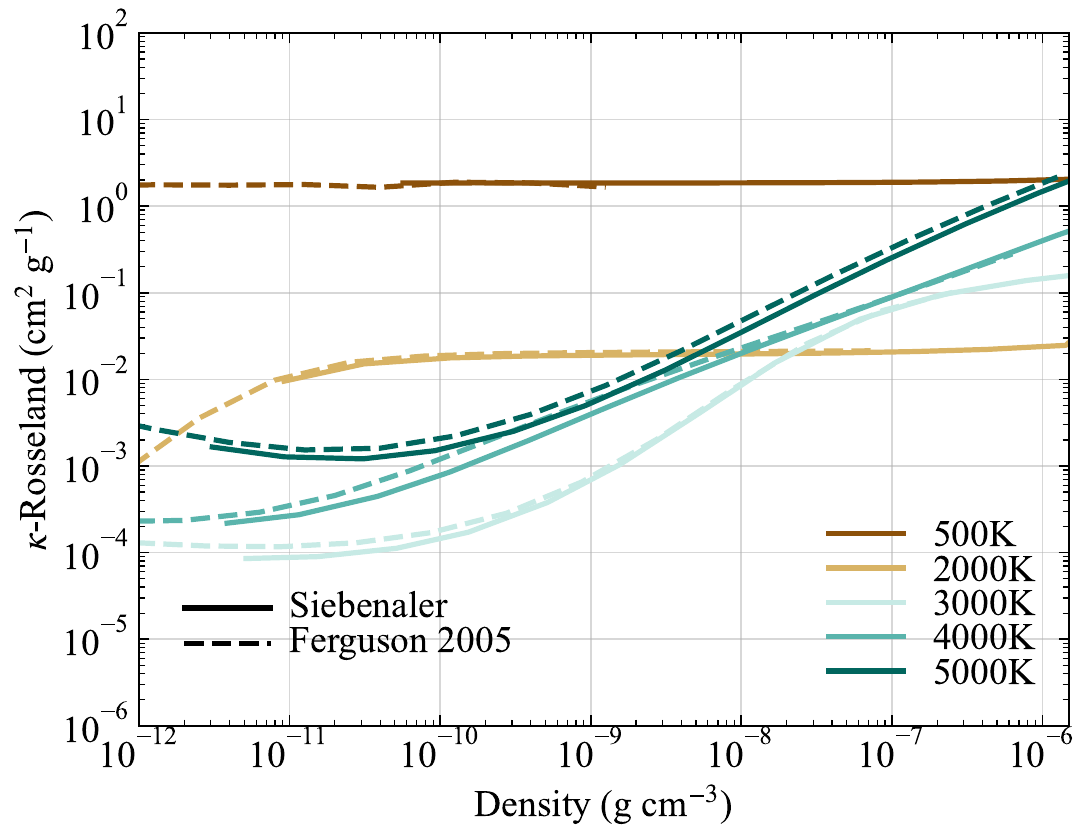}
    \caption{Comparison between our local mean opacities using equilibrium condensation and \citet{Ferguson_2005} including the contribution from grain opacities. Solid curves correspond to our calculations using $r_{\rm g} = 0.01 \ \rm \mu m$ and $\rm [M/H] = + 0.5$. Dashed curves were calculated by \citet{Ferguson_2005} (“ags04.7.04.tron” table) and assume $X = 0.70$ and $Z = 0.04$.}
   \label{fig:equ_cond_comp_Ferguson_2005}
\end{figure}

Fig. \ref{fig:equ_cond_Z+0.5} shows $\kappa_{\rm R}$ and $\kappa_{\rm P}$ at a pressure of  $1 \ \rm bar$ and $\textrm{[M/H]} = +0.5$ for the cloud-free (dashed curves) and cloudy (solid curves) cases. Since condensates do not settle into distinct layers, the opacity of cloud particles will increase $\kappa_{\rm R}$ and $\kappa_{\rm P}$ across all $T \lesssim 2000 \ \rm K$. In Fig. \ref{fig:equ_cond_comp_Ferguson_2005} we compare our $\kappa_{\rm R}$ values with those of \cite{Ferguson_2005} at fixed temperatures of 500, 2000, 3000, 4000, and 5000 K. We selected their table “ags04.7.04.tron”, which includes grain contributions and corresponds to a metallicity ($X = 0.70$, $Z = 0.04$) close to our $\rm [M/H] = +0.5$ case. The density overlap between the two data sets is limited since the \cite{Ferguson_2005} calculations are most appropriate for low density discs, while our data set applies to denser planetary atmospheres. However, we find that the agreement between the two data sets is generally good. The impact of grain opacities is only important in the $T = 500 \ \rm K$ case. \cite{Ferguson_2005} adopted a power-law distribution applicable to grains in the interstellar medium \citep{Mathis_1977}, while we assumed a log-normal distribution with $\sigma_{\rm g } = 2$ and $r_{\rm g} = 0.01 \ \rm \mu m$. Despite the differences in size distributions, the agreement is good.

\newpage
\section{Cross-section tables}
\input{molecules_table}

\input{CIA_table}
\input{clouds_table}
\input{FF_BF_table}


\section{Mean Opacity Tables}

\begin{table*} 
\centering
\caption{Cloud-free mean opacities for $\rm [M/H] = 0$.}
\label{table:cloud_free_table}
\setlength{\tabcolsep}{2.5pt} 
\footnotesize                   
\begin{tabular}{ccc cc cc cc cc cc cc}
\hline
 \hline
 &  &  
& \multicolumn{2}{c}{$T_{\rm g}$}
& \multicolumn{2}{c}{$T_{\rm eff}=3000$ K}
& \multicolumn{2}{c}{$T_{\rm eff}=4000$ K}
& \multicolumn{2}{c}{$T_{\rm eff}=5000$ K}
& \multicolumn{2}{c}{$T_{\rm eff}=6000$ K}
& \multicolumn{2}{c}{$T_{\rm eff}=7000$ K} \\
\cmidrule(lr){4-5} \cmidrule(lr){6-7} \cmidrule(lr){8-9} 
\cmidrule(lr){10-11} \cmidrule(lr){12-13} \cmidrule(lr){14-15}
$T$ & $P$ & $\rho_{\rm}$
& $\kappa_{\rm R}$ & $\kappa_{\rm P}$
& $\kappa_{\rm R}$ & $\kappa_{\rm P}$
& $\kappa_{\rm R}$ & $\kappa_{\rm P}$
& $\kappa_{\rm R}$ & $\kappa_{\rm P}$
& $\kappa_{\rm R}$ & $\kappa_{\rm P}$
& $\kappa_{\rm R}$ & $\kappa_{\rm P}$ \\
(K) & (bar) & (g cm$^{-3}$) & (cm$^2$ g$^{-1}$) & (cm$^2$ g$^{-1}$) & (cm$^2$ g$^{-1}$) & (cm$^2$ g$^{-1}$) & (cm$^2$ g$^{-1}$) & (cm$^2$ g$^{-1}$) & (cm$^2$ g$^{-1}$) & (cm$^2$ g$^{-1}$) & (cm$^2$ g$^{-1}$) & (cm$^2$ g$^{-1}$) & (cm$^2$ g$^{-1}$) & (cm$^2$ g$^{-1}$) 

 \\
100 & 1E-6 & 2.772E-10 & 2.119E-8 & 8.362E-2 & 1.322E-6 & 1.539E-1 & 2.915E-6 & 8.135E-2 & 5.449E-6 & 4.780E-2 & 9.131E-6 & 3.061E-2 & 1.417E-5 & 2.125E-2 
 \\
 100 & 3E-6 & 8.316E-10 & 6.343E-8 & 8.499E-2 & 1.627E-6 & 1.539E-1 & 3.555E-6 & 8.135E-2 & 6.602E-6 & 4.780E-2 & 1.101E-5 & 3.061E-2 & 1.703E-5 & 2.125E-2
 \\
 100 & 1E-5 & 2.772E-9 & 2.104E-7 & 8.976E-2 & 2.047E-6 & 1.539E-1 & 4.431E-6 & 8.135E-2 & 8.171E-6 & 4.779E-2 & 1.355E-5 & 3.061E-2 & 2.087E-5 & 2.125E-2
 \\
 ... & ... & ... & ... & ... & ... & ... & ... & ... & ... & ... & ... & ... & ... & ... 
 \\
 ... & ... & ... & ... & ... & ... & ... & ... & ... & ... & ... & ... & ... & ... & ...  
 \\
 100 & 3E0 & 8.311E-4 & 3.697E-2 & 7.926E-2 & 7.108E-5 & 1.451E-1 & 9.551E-5 & 7.745E-2 & 1.258E-4 & 4.574E-2 & 1.647E-4 & 2.939E-2 & 2.136E-4 & 2.047E-2
 \\
 100 & 1E1 & 2.770E-3 & 1.232E-1 & 2.640E-1 & 1.131E-4 & 1.709E-1 & 1.338E-4 & 9.285E-2 & 1.637E-4 & 5.538E-2 & 2.050E-4 & 3.575E-2 & 2.583E-4 & 2.486E-2
 \\
 100 & 3E1 & 8.311E-3 & 3.694E-1 & 7.920E-1 & 1.683E-4 & 2.449E-1 & 1.735E-4 & 1.369E-1 & 1.987E-4 & 8.295E-2 & 2.398E-4 & 5.395E-2 & 2.953E-4 & 3.742E-2
 \\
 ... & ... & ... & ... & ... & ... & ... & ... & ... & ... & ... & ... & ... & ... & ...  
 \\
 ... & ... & ... & ... & ... & ... & ... & ... & ... & ... & ... & ... & ... & ... & ... 
 \\
 1000 & 1E-6 & 2.783E-11 & 5.358E-5 & 7.205E-1 & 6.809E-5 & 8.646E0 & 8.432E-5 & 1.954E1 & 1.102E-4 & 2.663E1 & 1.458E-4 & 2.898E1 & 1.919E-4 & 2.828E1
 \\
 1000 & 3E-6 & 8.348E-11 & 5.628E-5 & 7.204E-1 & 6.841E-5 & 8.475E0 & 8.459E-5 & 1.920E1 & 1.104E-4 & 2.618E1 & 1.459E-4 & 2.849E1 & 1.918E-4 & 2.778E1
 \\
 1000 & 1E-5 & 2.783E-10 & 6.041E-5 & 7.201E-1 & 6.898E-5 & 8.271E0 & 8.514E-5 & 1.878E1 & 1.110E-4 & 2.561E1 & 1.465E-4 & 2.785E1 & 1.923E-4 & 2.712E1
 \\
 ... & ... & ... & ... & ... & ... & ... & ... & ... & ... & ... & ... & ... & ... & ... 
 \\
 ... & ... & ... & ... & ... & ... & ... & ... & ... & ... & ... & ... & ... & ... & ... 
 \\
 1000 & 3E4 & 8.360E-1 & 1.804E1 & 1.318E2 & 1.187E-2 & 5.628E1 & 4.406E-3 & 3.296E1 & 2.798E-3 & 2.046E1 & 2.305E-3 & 1.343E1 & 2.177E-3 & 9.245E0
 \\
 1000 & 5E4 & 1.393E0 & 2.807E1 & 2.183E2 & 8.419E-3 & 9.352E1 & 3.277E-3 & 5.479E1 & 2.171E-3 & 3.401E1 & 1.846E-3 & 2.233E1 & 1.786E-3 & 1.537E1
 \\
 1000 & 1E5 & 2.787E0 & 4.633E1 & 4.345E2 & 3.490E-3 & 1.866E2 & 1.505E-3 & 1.094E2 & 1.096E-3 & 6.789E1 & 1.005E-3 & 4.457E1 & 1.030E-3 & 3.068E1
 \\
 ... & ... & ... & ... & ... & ... & ... & ... & ... & ... & ... & ... & ... & ... & ... 
 \\
 ... & ... & ... & ... & ... & ... & ... & ... & ... & ... & ... & ... & ... & ... & ... 
 \\
 6000 & 1E5 & 4.552E-1 & 5.869E1 & 2.212E2 & 8.832E1 & 6.139E2 & 6.847E1 & 3.734E2 & 6.175E1 & 2.643E2 & 5.869E1 & 2.212E2 & 5.542E1 & 2.119E2
 \\
\hline

\end{tabular}
\end{table*}

\begin{table*}
\centering
\caption{Cloudy mean opacities for $\rm [M/H] = 0$ with a mean cloud particle radius $r_{\rm g} = 1 \ \rm \mu m$. 
}
\label{table:cloudy_table}
\setlength{\tabcolsep}{2.5pt} 
\footnotesize                   
\begin{tabular}{ccc cc cc cc cc cc cc}
\hline
 \hline
 &  &  
& \multicolumn{2}{c}{$T_{\rm g}$}
& \multicolumn{2}{c}{$T_{\rm eff}=3000$ K}
& \multicolumn{2}{c}{$T_{\rm eff}=4000$ K}
& \multicolumn{2}{c}{$T_{\rm eff}=5000$ K}
& \multicolumn{2}{c}{$T_{\rm eff}=6000$ K}
& \multicolumn{2}{c}{$T_{\rm eff}=7000$ K} \\
\cmidrule(lr){4-5} \cmidrule(lr){6-7} \cmidrule(lr){8-9} 
\cmidrule(lr){10-11} \cmidrule(lr){12-13} \cmidrule(lr){14-15}
$T$ & $P$ & $\rho_{\rm}$
& $\kappa_{\rm R}$ & $\kappa_{\rm P}$
& $\kappa_{\rm R}$ & $\kappa_{\rm P}$
& $\kappa_{\rm R}$ & $\kappa_{\rm P}$
& $\kappa_{\rm R}$ & $\kappa_{\rm P}$
& $\kappa_{\rm R}$ & $\kappa_{\rm P}$
& $\kappa_{\rm R}$ & $\kappa_{\rm P}$ \\
(K) & (bar) & (g cm$^{-3}$) & (cm$^2$ g$^{-1}$) & (cm$^2$ g$^{-1}$) & (cm$^2$ g$^{-1}$) & (cm$^2$ g$^{-1}$) & (cm$^2$ g$^{-1}$) & (cm$^2$ g$^{-1}$) & (cm$^2$ g$^{-1}$) & (cm$^2$ g$^{-1}$) & (cm$^2$ g$^{-1}$) & (cm$^2$ g$^{-1}$) & (cm$^2$ g$^{-1}$) & (cm$^2$ g$^{-1}$) 

 \\

 100 & 1E-6 & 2.772E-10 & 8.246E-5 & 8.388E-2 & 1.314E-3 & 1.563E-1 & 1.245E-3 & 8.305E-2 & 1.293E-3 & 4.917E-2 & 1.413E-3 & 3.181E-2 & 1.587E-3 & 2.235E-2 
 \\
 100 & 3E-6 & 8.316E-10 & 3.007E-5 & 8.505E-2 & 4.974E-4 & 1.544E-1 & 5.074E-4 & 8.175E-2 & 5.614E-4 & 4.816E-2 & 6.482E-4 & 3.095E-2 & 7.639E-4 & 2.157E-2
 \\
 100 & 1E-5 & 2.772E-9 & 1.176E-5 & 8.978E-2 & 1.863E-4 & 1.540E-1 & 2.153E-4 & 8.145E-2 & 2.596E-4 & 4.789E-2 & 3.198E-4 & 3.070E-2 & 3.964E-4 & 2.134E-2
 \\
 ... & ... & ... & ... & ... & ... & ... & ... & ... & ... & ... & ... & ... & ... & ... 
 \\
 ... & ... & ... & ... & ... & ... & ... & ... & ... & ... & ... & ... & ... & ... & ...  
 \\
 100 & 3E0 & 8.311E-4 & 4.152E-2 & 9.034E-2 & 1.100E-4 & 1.738E-1 & 1.224E-4 & 9.419E-2 & 1.488E-4 & 5.610E-2 & 1.871E-4 & 3.618E-2 & 2.371E-4 & 2.514E-2 
 \\
 100 & 1E1 & 2.770E-3 & 1.248E-1 & 2.674E-1 & 1.516E-4 & 1.795E-1 & 1.553E-4 & 9.787E-2 & 1.801E-4 & 5.848E-2 & 2.198E-4 & 3.779E-2 & 2.731E-4 & 2.626E-2
 \\
 100 & 3E1 & 8.311E-3 & 3.700E-1 & 7.931E-1 & 1.965E-4 & 2.477E-1 & 1.862E-4 & 1.386E-1 & 2.074E-4 & 8.399E-2 & 2.471E-4 & 5.462E-2 & 3.024E-4 & 3.789E-2
 \\
 ... & ... & ... & ... & ... & ... & ... & ... & ... & ... & ... & ... & ... & ... & ...  
 \\
 ... & ... & ... & ... & ... & ... & ... & ... & ... & ... & ... & ... & ... & ... & ... 
 \\
 1000 & 1E-6 & 2.783E-11 & 1.080E-1 & 8.228E-1 & 1.077E-1 & 8.754E0 & 1.038E-1 & 1.964E1 & 1.016E-1 & 2.673E1 & 1.006E-1 & 2.908E1 & 1.005E-1 & 2.838E1
 \\
 1000 & 3E-6 & 8.348E-11 & 4.563E-2 & 7.613E-1 & 4.284E-2 & 8.517E0 & 4.118E-2 & 1.924E1 & 4.044E-2 & 2.622E1 & 4.036E-2 & 2.852E1 & 4.078E-2 & 2.782E1
 \\
 1000 & 1E-5 & 2.783E-10 & 3.379E-2 & 7.502E-1 & 2.931E-2 & 8.300E0 & 2.817E-2 & 1.881E1 & 2.776E-2 & 2.564E1 & 2.785E-2 & 2.787E1 & 2.835E-2 & 2.715E1
 \\
 ... & ... & ... & ... & ... & ... & ... & ... & ... & ... & ... & ... & ... & ... & ... 
 \\
 ... & ... & ... & ... & ... & ... & ... & ... & ... & ... & ... & ... & ... & ... & ... 
 \\
 1000 & 3E4 & 8.360E-1 & 1.917E1 & 1.318E2 & 4.904E-2 & 5.628E1 & 2.307E-2 & 3.297E1 & 1.600E-2 & 2.047E1 & 1.336E-2 & 1.344E1 & 1.233E-2 & 9.252E0
 \\
 1000 & 5E4 & 1.393E0 & 3.002E1 & 2.183E2 & 3.345E-2 & 9.352E1 & 1.496E-2 & 5.479E1 & 1.030E-2 & 3.402E1 & 8.664E-3 & 2.233E1 & 8.105E-3 & 1.537E1
 \\
 1000 & 1E5 & 2.787E0 & 5.478E1 & 4.345E2 & 1.804E-2 & 1.866E2 & 7.812E-3 & 1.094E2 & 5.431E-3 & 6.789E1 & 4.676E-3 & 4.457E1 & 4.493E-3 & 3.068E1
 \\
 ... & ... & ... & ... & ... & ... & ... & ... & ... & ... & ... & ... & ... & ... & ... 
 \\
 ... & ... & ... & ... & ... & ... & ... & ... & ... & ... & ... & ... & ... & ... & ... 
 \\
 6000 & 1E5 & 4.552E-1 & 5.869E1 & 2.212E2 & 8.832E1 & 6.139E2 & 6.847E1 & 3.734E2 & 6.175E1 & 2.643E2 & 5.869E1 & 2.212E2 & 5.542E1 & 2.119E2
\\
\hline

\end{tabular}
\end{table*}


\bsp	
\label{lastpage}
\end{document}

%% file: molecules_table.tex
\begin{table*}
\small
\centering
 \caption{Molecular opacities used in this work.}
 \begin{tabular}[!b]{c @{\hskip 0.8cm } c @{\hskip 0.8cm } c @{\hskip 1.5cm } c @{\hskip 1.3cm } c @{\hskip 0.3cm } c} 
 \hline
 \hline
Molecule  & $T_{\rm max}$ (K) & Wavelength ($\mu \rm m$) & Line List Name &  References \\ [0.01ex] 
  \noalign{\smallskip}
    \hline
    \noalign{\smallskip}
    AlH & $5000$ & $0.407 - 500$ & AloHa &\textrm{\cite{Sergei_2023}}  \\
    CaH & $5000$ & $0.335 - 500$ & XAB &\textrm{\cite{Owens_2022a}}  \\
    CaO & $6000$ & $0.400 - 500$ & VBATHY &\textrm{\cite{Yurchenko_2016}}  \\
    CaOH & $5000$ & $0.278 - 500$ & OYT6 &\textrm{\cite{Owens_2022}}  \\
    CH & $6000$ & $0.255 - 200$ & MoLLIST &\textrm{\cite{Masseron_2014, Bernath_2020}}  \\
    CH$_4$ & $5000$ & $0.833 - 500$ & MM & \textrm{\cite{Yurchenko_2024}}  \\
    CO & $900$ & $0.455 - 500$ & Li2015 &\textrm{\cite{Li_2015, Somogyi_2021}}  \\
    CO$_2$ & $5000$ & $0.500 - 500$ & UCL-4000 &\textrm{\cite{Yurchenko_2020}}  \\
    CP & $3000$  & $0.661 - 28$ & MoLLIST &[1] \\
    CrH & $3000$  & $0.667 - 1.615$ & MoLLIST & [2]  \\
    FeH & $6000$  & $0.667 - 50$ & MoLLIST &\textrm{\cite{Dulick_2003, Bernath_2020}} \\
    H$_2$ & $6000$ & $0.278 - 200$ & RACPPK &\textrm{\cite{Roueff_2019}}\\
    H$_2$O & $6000$ & $0.243 - 500$ & POKAZATEL &\textrm{\cite{Polyansky_2018}}\\
     H$_2$S & $3000$ & $0.286 - 500$ & AYT2 &\textrm{\cite{Azzam_2016, Chubb_2018}}\\
    HCl & $5000$ & $0.494 - 500$ & HITRAN-HCl &\textrm{\cite{Gordon_2017}}\\
    HCN & $4000$ & $0.569 - 500$ & Harris &\textrm{\cite{Harris_2006, Barber_2013}}\\
    HF & $5000$ & $0.31 - 500$ & Coxon-Hajig & [3]\\
    LiOH & $5000$ & $1 - 500$ & OYT7 &\textrm{\cite{Owens_2024}}\\
    MgH & $5000$ & $0.338 - 500$ & XAB &\textrm{\cite{Owens_2022a}}\\
    MgO & $5000$ & $0.270 - 500$ & LiTY &\textrm{\cite{Li_2019}}\\
    N$_2$ & $6000$ & $0.179 - 500$ & WCCRMT & [4]\\
    NaCl & $3000$  & $4.069 - 500$ & Barton &\textrm{\cite{Barton_2014}}\\
    NaH & $6000$ & $0.311 - 500$ & Rivlin &\textrm{\cite{Rivlin_2015, Chubb_2020}}\\
    NH$_3$ & $2000$ & $0.500 - 500$ & CoYuTe &\textrm{\cite{Derzi_2015, Coles_2019}}\\
    PH$_3$ & $3000$ & $1 - 500$ & SAITY &\textrm{\cite{Silva_2014}}\\
    PN & $5000$ & $0.121 - 500$ & PaiN &\textrm{\cite{Semenov_2024}}\\
    PS & $5000$ & $0.270 - 500$ & POPS &\textrm{\cite{Prajapat_2017}}\\
    SiH & $5000$ & $0.313 - 500$ & SiGHTLY &\textrm{\cite{Yurchenko_2017}}\\
    SiH$_4$ & $2000$ & $2 - 500$ & OY2T &\textrm{\cite{Owens_2017}}\\
    SiO & $6000$  & $0.139 - 500$ & SiOUVenIR &\textrm{\cite{Yurchenko_2021}}\\
    SO & $5000$ & $0.222 - 500$ & SOLIS &\textrm{\cite{Brady_2023}}\\
    TiH & $4800$ & $0.417 - 2.156$ & MoLLIST &\textrm{\cite{Burrows_2005, Bernath_2020}}\\
    TiO & $6000$ & $0.333 - 500$ & Toto &\textrm{\cite{McKemmish_2019}}\\
    VO & $5400$ & $0.222 - 500$ & HyVO &\textrm{\cite{Bowesman_2024}}\\
        
  \hline
\end{tabular}
 \label{table:molecule_opacity}
 \begin{flushleft}
\tiny{[1]: \textrm{\cite{Ram_2014, Bernath_2020, Qin_2021}} [2]: \textrm{\cite{Burrows_2002, Chubb_2018, Bernath_2020}} [3]: \textrm{\cite{Li_2015, Coxon_2015, Somogyi_2021}} [4]: \textrm{\cite{Shemansky_1969, Western_2017, Western_2018, Jans_2024}}} 
\end{flushleft}
 
\end{table*}

%% file: CIA_table.tex

\begin{table*} 
\small
\centering
 \caption{Collision-induced absorption used in this work.}
 \begin{tabular}[t]{c @{\hskip 1.25cm }  c @{\hskip 1.25cm } c @{\hskip 1.25cm } c @{\hskip 0.1cm } c} 
 \hline
 \hline
Species  & Temperature range & Wavelength ($\mu \rm m$) & References \\ [0.01ex] 
  \noalign{\smallskip}
    \hline
    \noalign{\smallskip}
    H$_2$-H$_2$ & $100 - 400$  & $0.5 - 500$ & \cite{Borysow_2002, Fletcher_2018, Orton_2025}  \\
    & $400 - 3000$ &  & \cite{Borysow_2001, Borysow_2002, Abel_2012}
    \\
    & $3000 - 5000$ &  & \cite{Borysow_2001}
    \\
    H$_2$-He & $100 - 200$ & $0.5 - 500$ & \cite{Borysow_1989, Borysow_1989_2, Orton_2025} \\
    & $200 - 6000$ & & \cite{Abel_2011}
    \\
    H$_2$-H & $1000 - 2500$ & $1 - 100$ & \textrm{\cite{Gustafsson_2003}} \\
    H$_2$-CH$_4$ & $100 - 400$ &  $5.139 - 500$ & \textrm{\cite{Borysow_1986}}  \\
    H$_2$-CO$_2$ & $200 - 350$ &  $5 - 500$ & \textrm{\cite{Wordsworth_2017}}  \\
    He-H & $1500 - 6000$ &  $0.9 - 200$ & \textrm{\cite{Gustafsson_2001}}  \\
    He-CH$_4$ & $100 - 350$ &  $10 - 500$ & \textrm{\cite{Taylor_1988}}  \\
    CH$_4$-CH$_4$ & $100 - 400$ &  $10.1 - 500$ & \textrm{\cite{Borysow_1987}}  \\

  \hline
\end{tabular}
 \label{table:CIA_opacity}

\end{table*}

%% file: clouds_table.tex
\begin{table*}
\small
\centering
 \caption{Grain/cloud opacities considered in this work.}
 \begin{tabular}{c @{\hskip 1.8cm }  c @{\hskip 1.8cm } c @{\hskip 1.8cm }  c @{\hskip 0.1cm } c} 
 \hline
 \hline
Condensate & Wavelength ($\mu$m) & References \\ [0.01ex] 
  \noalign{\smallskip}
    \hline
    \noalign{\smallskip}
    Al$_2$O$_3$ & $0.2 - 500$ &   \textrm{\cite{Koike_1995, Begemann_1997}}  \\
    Ca$_2$Al$_2$SiO$_7$ & $6.690 - 500$ &   \textrm{\cite{Mutschke_1998}}  \\
    Ca$_2$SiO$_4$ & $0.196 - 500$ &   \textrm{\cite{Jäger_2003}}  \\
    CaTiO$_3$ & $0.1 - 500$ &   \textrm{\cite{ Ueda_1998, Posch_2003}}  \\
    Cr & $0.1 - 500$ &   \textrm{\cite{Palik_1991, Rakic_1998}}  \\
    Cu & $0.517 - 5.560$ &   \textrm{\cite{Ordal_1985}}  \\
    Fe & $0.1 - 285.7$ &   \textrm{\cite{Palik_1991}} \\
    Fe$_2$O$_3$ & $0.1 - 500$ &   \href{https://www.physik.uni-jena.de/19613/aiu}{A.H.M.J. Triaud, DOCCD Jena Laboratory}  \\
    FeO & $0.2 - 500$ &   \textrm{\cite{Henning_1995}} \\
    FeS & $0.1 - 487.381$ &   \textrm{\cite{Pollack_1994, Henning_1997}} \\
    H$_2$O & $0.1 - 500$ &   \textrm{\cite{Warren_1984}}  \\
    KCl & $0.1 - 487.381$ &   \textrm{\cite{Palik_1985}} \\
    Mg$_2$SiO$_4$ & $0.196 - 500$ &   \textrm{\cite{Jäger_2003}}  \\
    MgAl$_2$O$_4$ & $0.35 - 500$ &   \textrm{\cite{Palik_1991, Zeidler_2011}} \\
    MgO & $0.1 - 500$ &   \textrm{\cite{Palik_1991}} \\
    MgSiO$_3$ & $0.196 - 500$ &   \textrm{\cite{Palik_1991, Zeidler_2011}}
    \\
    MnS & $0.196 - 190$ &   \textrm{\cite{Huffman_1967, Montaner_1979}}
    \\
    Na$_2$S & $0.1 - 200$ &   \textrm{\cite{Montaner_1979, Khachai_2009}}
    \\
    NaAlSi$_3$O$_8$ & $6.699 - 500$ &   \textrm{\cite{Mutschke_1998}}
    \\
    NaCl & $0.1 - 500$ &   \textrm{\cite{Palik_1985}}
    \\
    NH$_3$ & $1.67 - 50$ &   \textrm{\cite{Trotta_1996, Hudson_2022}}
    \\
    NH$_4$SH & $2 - 20$ &   \textrm{\cite{Howett_2006, Gerakines_2024}}
    \\
    Ni & $0.667 - 286$ &   \textrm{\cite{Ordal_1987}}
    \\
    SiO & $0.1 - 100.858$ &   \textrm{\cite{Palik_1985}}
    \\
     SiO$_2$ & $0.1 - 500$ &   \textrm{\cite{Palik_1985, Zeidler_2013}}
    \\
    TiO$_2$ & $0.12 - 500$ &   \textrm{\cite{Posch_2003, Zeidler_2011, Siefke_2016}}
    \\

  \hline
\end{tabular}
 \label{table:cloud_opacity}

\end{table*}

%% file: FF_BF_table.tex
\begin{table*}
\small
\centering
 \caption{Free-free and bound-free absorptions considered in this work.}
 \begin{tabular}{c @{\hskip 1.8cm }  c @{\hskip 1.8cm } c @{\hskip 1.8cm }  c @{\hskip 0.1cm } c} 
 \hline
 \hline
Reaction & Wavelength ($\mu$m) & References \\ [0.01ex] 
  \noalign{\smallskip}
    \hline
    \noalign{\smallskip}
    $\textrm{H}_2 + \textrm{e}^- + \textrm{h}\nu \xrightarrow{} \textrm{H}_2 +  \textrm{e}^-$ & $0.351 - 500$ &  \textrm{\cite{Bell_1980}}  \\
     $\textrm{H} + \textrm{e}^- + \textrm{h}\nu \xrightarrow{} \textrm{H} +  \textrm{e}^-$ & $0.182 - 500$ &\textrm{\cite{John_1988}}  \\
     $\textrm{He} + \textrm{e}^- + \textrm{h}\nu \xrightarrow{} \textrm{He} +  \textrm{e}^-$ & $0.506 - 500$ & \textrm{\cite{John_1994}}  \\
     $\textrm{Li} + \textrm{e}^- + \textrm{h}\nu \xrightarrow{} \textrm{Li} +  \textrm{e}^-$ &$0.5 - 500$ & \textrm{\cite{John_1975}}  \\
     $\textrm{N} + \textrm{e}^- + \textrm{h}\nu \xrightarrow{} \textrm{N} +  \textrm{e}^-$ &  $0.5 - 500$ & \textrm{\cite{John_1975}}  \\
     $\textrm{O} + \textrm{e}^- + \textrm{h}\nu \xrightarrow{} \textrm{O} +  \textrm{e}^-$ &  $0.5 - 500$ & \textrm{\cite{John_1975}}  \\
     $\textrm{Na} + \textrm{e}^- + \textrm{h}\nu \xrightarrow{} \textrm{Na} +  \textrm{e}^-$ & $0.5 - 500$ & \textrm{\cite{John_1975}}  \\
     $\textrm{CO} + \textrm{e}^- + \textrm{h}\nu \xrightarrow{} \textrm{CO} +  \textrm{e}^-$ & $0.1 - 500$ & \textrm{\cite{John_1975}}  \\
     $\textrm{N}_2 + \textrm{e}^- + \textrm{h}\nu \xrightarrow{} \textrm{N}_2 +  \textrm{e}^-$ & $0.1 - 500$ & \textrm{\cite{John_1975}}  \\
     $\textrm{H}_2\textrm{O} + \textrm{e}^- + \textrm{h}\nu \xrightarrow{} \textrm{H}_2\textrm{O} +  \textrm{e}^-$ & $0.1 - 500$ & \textrm{\cite{John_1975}}  \\
     $\textrm{H}^-  + \textrm{h}\nu \xrightarrow{} \textrm{H} +  \textrm{e}^-$ & $0.1 - 1.644$ &  \textrm{\cite{McLaughlin_2017}}  \\
  \hline
\end{tabular}
 \label{table:BF_FF_opacity}

\end{table*}